\documentclass[aps,prstab, amsmath, amssymb, preprint, superscriptaddress, 12pt]{revtex4-1}
\usepackage{graphicx}
\usepackage{subfigure}

\begin{document}

\title{Quasi-monoenergetic femtosecond photon sources from Thomson Scattering using laser plasma accelerators and plasma channels}

\author{S.G.~Rykovanov}
\email{SRykovanov@lbl.gov}
\address{Lawrence Berkeley National Laboratory, Berkeley, CA 94720}

\author{C.G.R.~Geddes}
\email{CGRGeddes@lbl.gov}
\address{Lawrence Berkeley National Laboratory, Berkeley, CA 94720}

\author{J.-L. Vay}
\address{Lawrence Berkeley National Laboratory, Berkeley, CA 94720}

\author{C.B.~Schroeder}
\address{Lawrence Berkeley National Laboratory, Berkeley, CA 94720}

\author{E.~Esarey}
\address{Lawrence Berkeley National Laboratory, Berkeley, CA 94720}

\author{W.P.~Leemans}
\address{Lawrence Berkeley National Laboratory, Berkeley, CA 94720}

\date{\today}
\begin{abstract}
Narrow bandwidth, high energy photon sources can be generated by
Thomson scattering of laser light from energetic electrons, and
detailed control of the interaction is needed to produce high quality
sources.  We present analytic calculations of the energy-angular
spectra and photon yield that parametrize the influences of the
electron and laser beam parameters to allow source design.  These
calculations, combined with numerical simulations, are applied to
evaluate sources using conventional scattering in vacuum and methods
for improving the source via laser waveguides or plasma channels.  We
show that the photon flux can be greatly increased by using a plasma
channel to guide the laser during the interaction.  Conversely, we
show that to produce a given number of photons, the required laser
energy can be reduced by an order of magnitude through the use of a
plasma channel.  In addition, we show that a plasma can be used as a
compact beam dump, in which the electron beam is decelerated in a
short distance, thereby greatly reducing radiation shielding.
Realistic experimental errors such as transverse jitter are
quantitatively shown to be tolerable.  Examples of designs for sources
capable of performing nuclear resonance fluorescence and photofission
are provided.

\end{abstract}


\maketitle

\section{Introduction}
\label{intro}

Thomson Scattering (TS) of light from fast moving electrons is a
well-known and established source of X-ray and $\gamma$-ray radiation.
It was in the 1960s, after the discovery of the laser, when the first
TS x-ray sources were proposed \cite{Arutyunyan1963, Arutyunyan1964,
Milburn1963} and demonstrated in experiments~\cite{Kulikov1964}.
Since then many important results were obtained describing TS sources
\cite{Sprangle1992, Esarey1993, Ohgaki1994, Ting1995, Schoenlein1996, Leemans1996, Carroll1999, Pogorelsky2000, Pietralla2001, Catravas2001, Nakano2003,
Nedorezov2004, Leemans2005, Nedorezov2012,Achterhold2013}, including the first
demonstration of femtosecond X-ray pulses at the Accelerator Test
Facility of Lawrence Berkeley National Laboratory (LBNL)
\cite{Schoenlein1996, Leemans1996}.  Intense X- and $\gamma$-ray
sources can be used in many areas of science, industry and medicine.
Photons with energies above approximately 1 MeV serve as a probe for
nuclear physics (e.g, see the review by V.~Nedorezov \textit{et
al}~\cite{Nedorezov2004}).  An update on prospects for brilliant,
monochromatic $\gamma$-ray sources is given in
Ref.~\cite{Thirolf2012}.  Today, the HIGS facility at Duke
University~\cite{HIGS_Facility,Weller2009} is the most intense source
of narrow bandwidth (5 percent FWHM) $\gamma$-ray sources with photon
energies in the range from 1 to 100 MeV and with photon flux of about
$10^8$~photons/second, and is a large fixed facility.  Thomson
Scattering sources presently use large conventional particle
accelerators, including at LBNL~\cite{Schoenlein1996,Leemans1996},
HIGS facility at Duke University~\cite{HIGS_Facility,Weller2009},
TREX/MEGA-Ray facility at Lawrence Livermore National
Laboratory~\cite{Albert2010} and others~\cite{Ting1995,
Carroll1999,Pogorelsky2000a}.  For example, the HIGS facility produces
$~10^8$ photons per second with central energies $1-100$MeV and energy
spread of approximately 5$\%$.  Its planned update, HIGS2 is projected
to produce approximately same number of photons per second with
central energies $2-12$ MeV and FWHM energy spread $<0.5\%$.  A new
generation of sources is under construction including the Extreme
Light Infrastructure Nuclear Physics Facility~\cite{Thirolf2012}, a
project at the Japan Atomic Energy Agency~\cite{Ohgaki1994,Kawase2008,
Kawase2011}, and proposed facilities at Fermi National Accelerator
Laboratory~\cite{ASTA2013}, SLAC National Accelerator
Laboratory~\cite{FACETII} and Brookhaven National
Laboratory~\cite{ATF}.

Narrow bandwidth Thomson gamma ray sources are a powerful tool for  Nuclear Resonance Fluorescence (NRF), radiography and photo-fission studies for the detection of nuclear materials in cargo containers or nuclear waste (spent fuel)~\cite{Bertozzi2008,Albert2011a,Quiter2011a,Quiter2011,Thirolf2012,Tesileanu2013, Johnson2011}. 
Ability to produce narrow bandwidth intense $\gamma$-rays essentially defines screening time and radiation dose, and also the feasibility of industrial usage. NRF studies are the most demanding, and approximately $10^9- 10^{12}$ photons/sec in $\lesssim$ two percent bandwidth around the (element specific) NRF line is desired, with the range corresponding to different shielding from unshielded up to full cargo containers~\cite{Quiter2013, Hajima2009, Pruet2006}. For example, for $^{235}$U which is of primary interest for nuclear nonproliferation, the NRF line energy is around 1.7~MeV. In the case of photofission studies, similar fluxes are needed but with a more relaxed bandwidth of approximately 10 percent at energies from approximately 10 to 15 MeV is desired. Radiography is less demanding in both yield and energy spread. These requirements are outlined in Table~\ref{higs_table}. Compact, transportable sources capable of high fluxes and narrow bandwidth are needed.

\begin{center}
\begin{table}[b]
\caption{Requirements for the source capable of performing NRF studies of $^{235}$U and photofission.}
\begin{tabular}{lllcc}
\hline
  &  $^{235}$U NRF $\phantom{|}$& Photofission\\
\hline
$E_\gamma[\mathrm{MeV}]$     & 1.7 &  ~5-15   \\
$\Delta E_\gamma/E_\gamma$ [FWHM]          & $\sim2\%$ & $\sim10\%$    \\
Total Flux~[photons/sec] & $\sim10^8-10^9$ & $\sim10^9-10^{10}$  \\
Collimated Flux~[photons/sec] $\phantom{|}$   & $\sim 10^7$ & $\sim 10^8$\\
\hline
\end{tabular}
\label{higs_table}
\end{table}
\end{center}

Thomson scattering sources typically require much higher electron
energies than Bremsstrahlung sources to achieve a certain photon
energy increasing the size of the system.  With the rapid development
of accelerator and laser technology in the recent decades it has now
become possible to build dedicated TS machines.  For example, the
Compact Light Source developed by Lynceantech~\cite{Lynceantech} and
working in the keV hard x-ray regime can fit into the typical
laboratory.  It makes use of a storage ring for acceleration of
electrons.  Generation of MeV-level gamma-rays with storage rings or
linacs however leads to large accelerator size (approximately 20 to 50
meters) due to limitations in the accelerating
gradient~\cite{Hajima2009}.  Additionally, the low conversion factor
of scattering laser photons into X- or gamma-ray photons due to the
very small cross-section of the process ($\approx$0.7 b) requires
large scattering lasers.  These challenges limit current TS
applications, especially those requiring transportability.

\begin{figure}
   \centering
    \includegraphics[width=0.5\textwidth,trim=0cm 6.5cm 0cm 1.5cm, clip=true]{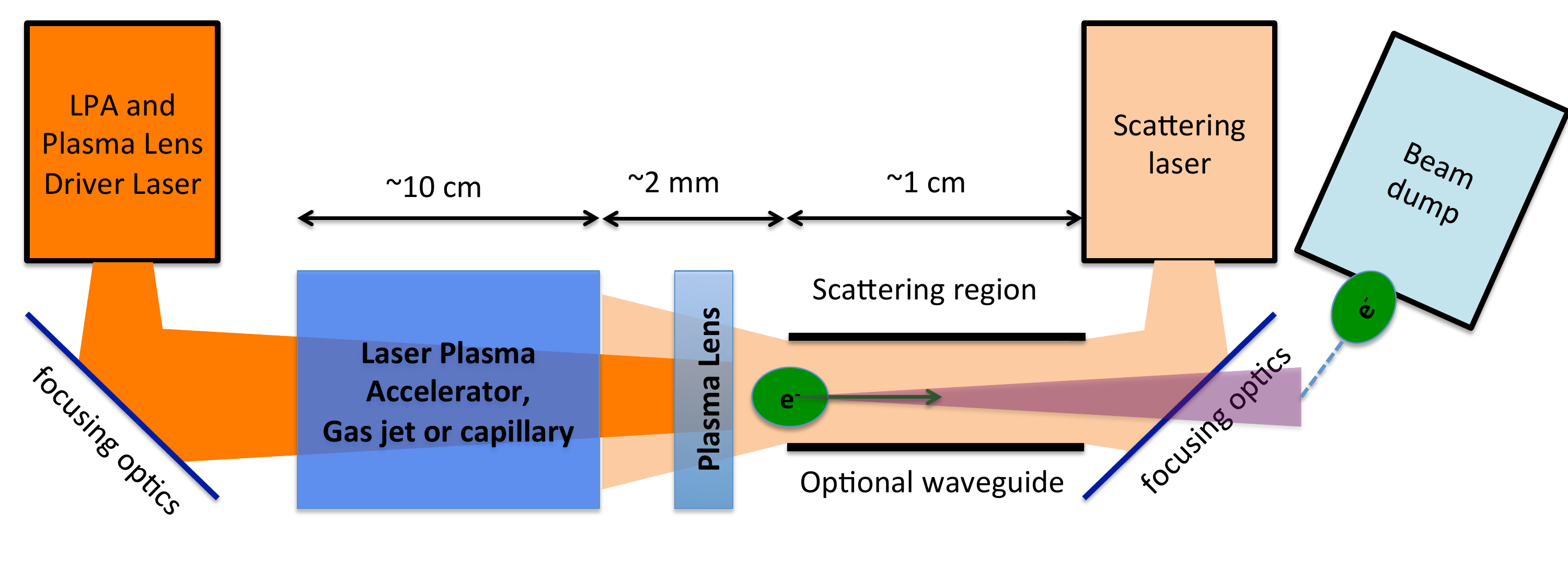}
     \caption{Conceptual source setup (not in scale) including the plasma lens and plasma channel waveguide.}\label{tentative_setup_figure}
\end{figure}

Recent advances in laser plasma accelerators (LPA)~\cite{Esarey2009}, 
where stable GeV-level electron beams have been produced in just 3 centimeters of acceleration 
distance~\cite{Leemans2006} allows one to consider compact Thomson sources. These
LPA electron energies in principle allows generation of hard photons
with energies up to 25~MeV (or up to 50~MeV with laser frequency
doubling).  Such photon energies are relevant for many applications
including photo-nuclear experiments \cite{Quiter2011a,Quiter2011},
ultrafast radiography \cite{Tommasini2011, Toyokawa2002} and cancer
therapy \cite{Weeks1997a}.  The intrinsic short duration of LPA
electron beams (on the order of several
femtoseconds~\cite{VanTilborg2006,Lundh2011}) also leads to an easy
setup for generation of femtosecond x-rays that can be useful in
time-resolved studies.  However, only broadband (i.e. with bandwidth
more than 20 percent) X- and gamma-ray sources have so far been
demonstrated using LPAs~\cite{Schwoerer2006,TaPhuoc2012,Chen2013a,
Powers2013}.  Designs which account for and utilize the unique
properties of LPA beams, and provide efficient scattering, are hence
needed to enable narrow bandwidth compact sources.

High flux Thomson sources require either large scattering lasers, or high electron beam current, or
novel solutions to increase effectiveness.  Indeed, for LPAs GeV
electron beams were produced using a 40~TW laser system, the size of
which have rapidly decreased and is now at the level of
6~m$^2$~\cite{ThalesLaser}.  Hence, the goal of compact source
development is for the scattering laser occupy approximately the same
area and not much more, while at the same time maximizing photon yield per electron. The later is important for compact sources where high electron current, which increases shielding needs, is undesirable. Increasing yield conventionally requires
laser pulses and electrons be tightly focused, but this approach is
limited.  One needs keep intensity low in order to avoid nonlinear
broadening effects, which in turn requires long pulse durations.
Similarly, as the interaction length is approximately equal to double
the Rayleigh range $Z_R$, focusing too tightly reduces $Z_R$ and hence
the total photon yield.  For a given laser pulse energy there then
exist an optimum laser pulse spot size and laser pulse duration to
maximize yield (depending on the interaction geometry).  The result is
that high scattering laser energies (much more that the LPA driving
laser) are required, dominating the total size of the source.  A
straightforward optimization strategy is to increase the interaction
length by using waveguides for diffractionless propagation of the
laser pulses.  Standard waveguides, such as, for example, metallic
tubes or hollow-core fibers can be in theory
used~\cite{Karagodsky2011,Plettner2008}, but will be destroyed quickly
and will need to be replaced frequently in experiments.  Pogorelsky
\textit{et al} \cite{Pogorelsky2000} proposed to use plasma
channels~\cite{III1993, Leemans2006, Geddes2004} for guiding CO$_2$
laser pulses for increasing the Thomson scattering yield, and this
concept must be further developed to design compact sources.

In this paper we study TS from LPAs, a simplified schematic of which
is presented on Fig.~\ref{tentative_setup_figure}.  We show that the
performance of LPA-based TS sources can be enhanced using plasma
channels.  We present analytic formulae and numerical considerations
for the spectral shape of the radiation taking into account realistic
laser pulses and electron beams as well as the total yield of TS
sources for different interaction geometries.  We provide examples of
compact TS sources based on LPA electrons, and quantitatively
demonstrate that experimental errors, such as, for example, transverse
jitter are tolerable within current laser and LPA technology.  We
demonstrate both analytically and numerically efficient TS source
designs using waveguides and plasma channels for control of laser and
electron beam propagation that may lead to considerable reduction in
size and cost of future sources.  We also study the use of a plasma as
a compact beam dump for the high energy electron beam.

The paper is organized as follows.  First, the basic mechanism of
photon generation using Thomson scattering from electron beams is
reviewed in Sec.~\ref{thomson_basics_section}.
Section~\ref{spectral_shape_section} is devoted to the spectral shape
of the generated radiation and effects that lead to broadening, such
as electron beam divergence (Sec.~\ref{spectral_shape_div_section}),
electron beam energy spread
(Sec.~\ref{spectral_shape_energy_spread_section}), laser pulse
intensity and multiple scattering
(Sec.~\ref{spectral_shape_intensity_section}).  In
Sec.~\ref{estimation_collimation_section} estimation of the
collimation angle and relative photon number for a given source
bandwidth are calculated.  In Sec.~\ref{using_LPA_section} a
discussion on using the LPA electron beams for generation of narrow
bandwidth X- and $\gamma$-ray sources is provided and required
electron beam manipulations are outlined.  In the same section a
compact LPA based beam dump is discussed.  In Sec.~\ref{yield_section}
derivation of the total photon yield for different interaction
geometries (vacuum, laser waveguide, plasma channel for guiding both
electron and laser beams) is presented.  In the same section yield
degradation due to the pointing errors is quantitatively evaluated.
In Sec.~\ref{numerical_sims_section} we present examples of design
calculations and numerical simulations of the realistic $\gamma$-ray
sources capable of performing the NRF studies of $^{235}$U and
photofission experiments.  Finally, Sec.~\ref{conclusions_section}
contains conclusions and final discussions.

\section{Basics of Thomson scattering}\label{thomson_basics_section}
\begin{figure}
   \centering
    \includegraphics[width=0.5\textwidth,trim=0.5cm 1.5cm 0cm 0.2cm, clip=true]{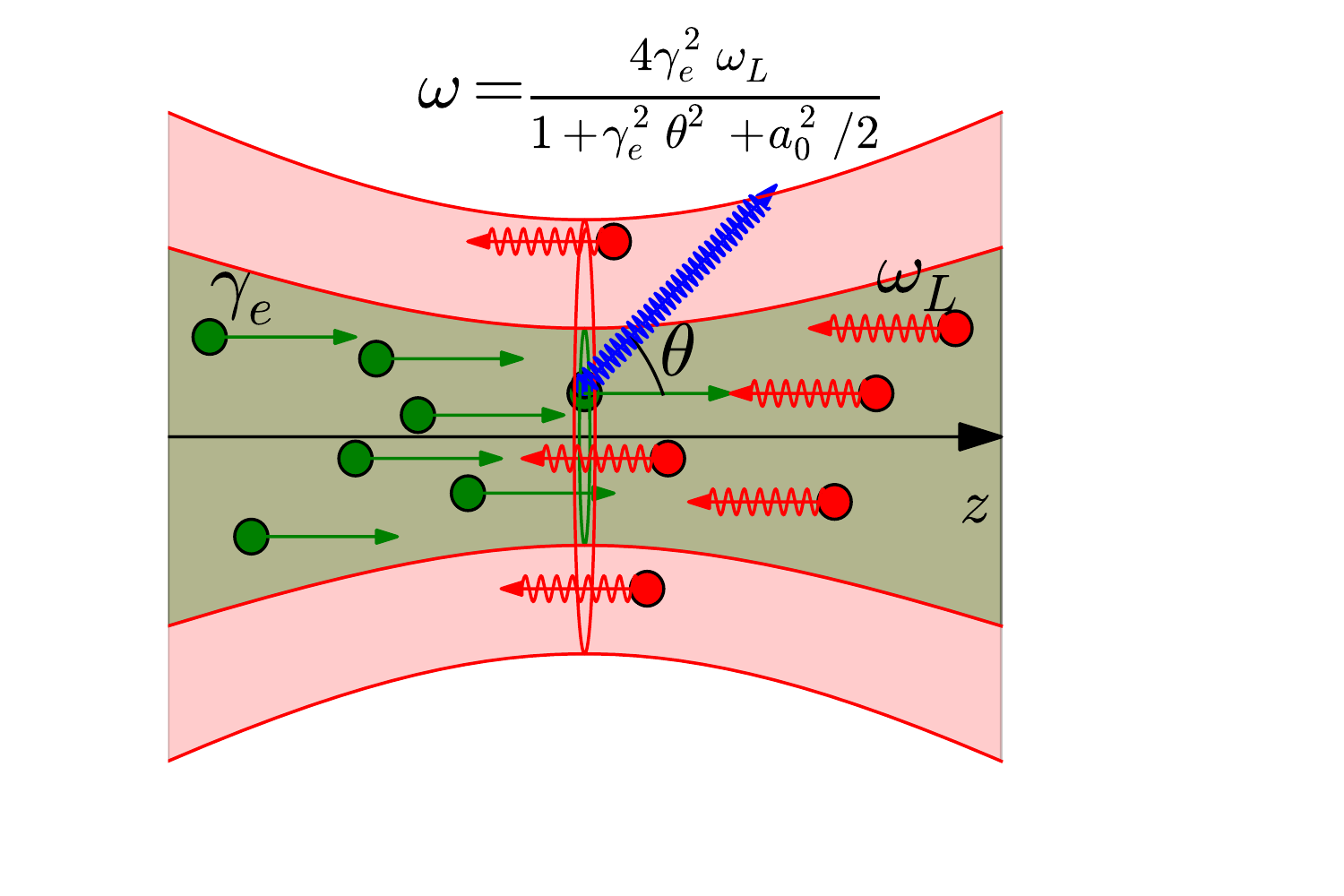}
     \caption{Schematic of a TS source in head-on geometry. Red circles represent laser photons with frequency $\omega_L$ and the red shaded area depicts the laser beam envelope undergoing focusing. Green circles and green shaded area represent electrons with energy $m_ec^2\gamma_e$ and the electron beam envelope respectively. A random act of scattering under the angle $\theta$ is shown in blue color.}\label{schematics_figure}
\end{figure}

\begin{figure}
   \centering
    \includegraphics[width=0.5\textwidth]{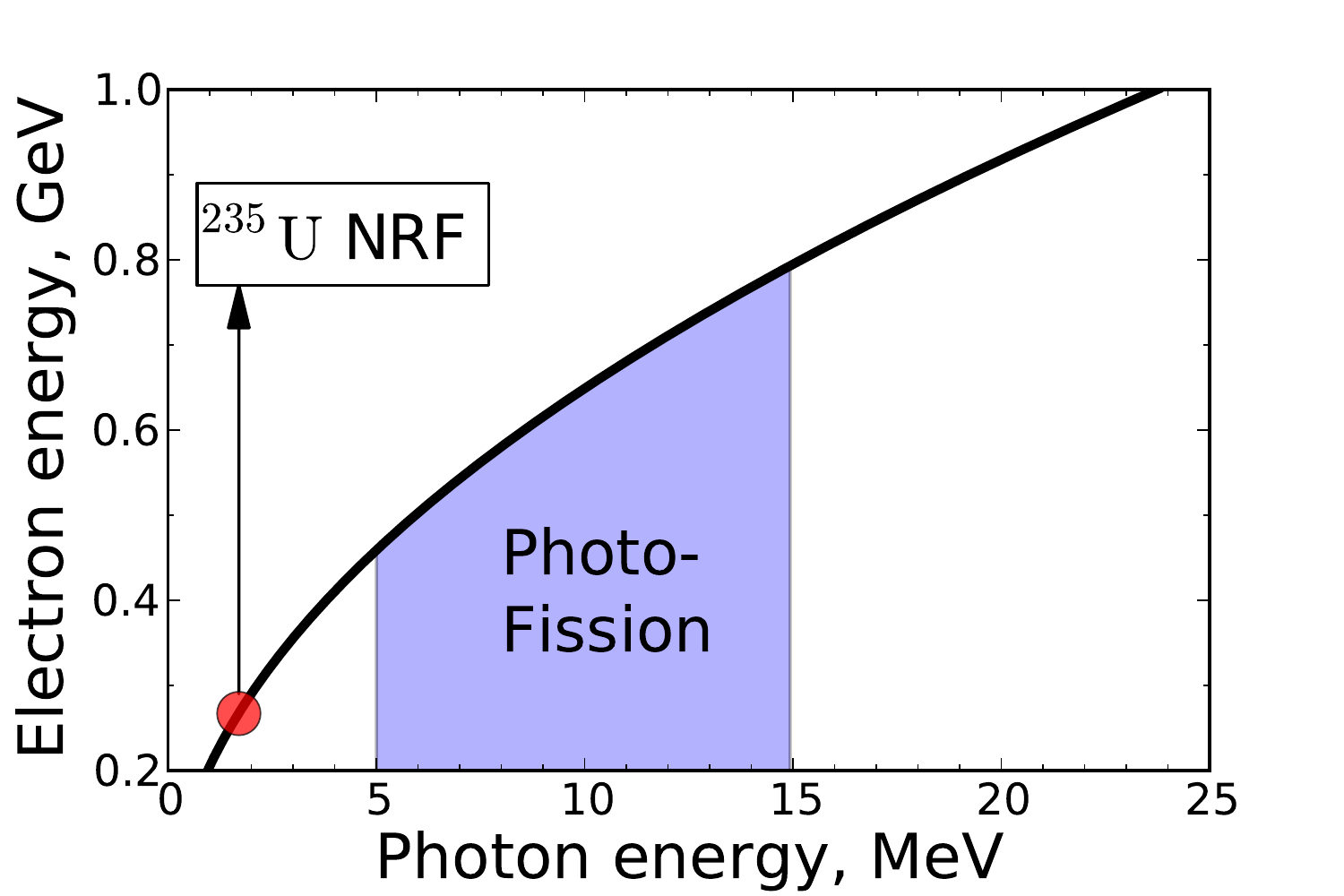}
     \caption{Electron energy (vertical axis) required for the generation of a photon with a certain energy (horizontal axis) according to eq.~(\ref{photon_energy_eq}) with $\theta=0$, and $\lambda_L=0.8\mu m$.}\label{photon_energy_figure}
\end{figure}

Thomson scattering, which includes also undulator radiation, is a well studied area of physics~\cite{Schmueser2008, Sarachik1970, Esarey1993, Catravas2001, Sprangle1992, Litvinenko1996}. A schematic of the TS source under consideration  is shown in Fig.~\ref{schematics_figure}. Maximum photon energy is obtained in the case when the laser photon and electron collide head-on and the photon is scattered exactly backwards (in other words at 180$^\circ$). In such a situation, assuming that the laser pulse is weak, photon energy is given by the well-known (double) relativistic Doppler shift formula $\hbar \omega=4\gamma_e^2\hbar \omega_L$, where $\omega$ is the generated photon frequency, $\omega_L$ is the laser frequency and $\gamma_e$ is the electron relativistic gamma-factor. Figure~\ref{photon_energy_figure} shows the plot of electron energy required for the generation of a photon with a certain energy according to this formula for a laser pulse with the commonly used wavelength $\lambda_L=0.8\mu m$. One can see that $\gamma$-ray energies required for the NRF studies (for example, the NRF line is around 1.7 MeV for $^{235}$U), photo-fission studies (requiring photon energies in the range from 5 to 15 MeV), radiography (broad energy range), and cancer treatment (MeV-level photons) are well within reach for LPA electrons, typically in the range from 0.1 to 1 GeV. Consider electrons  with energy $E=m_ec^2\gamma_e$ on the order of 0.1 to 1 GeV are traveling along the $z$-axis and collide head-on with the laser beam. Here $m_e$ and $c$ are electron mass and the speed of light in vacuum respectively. Photon energy in the case of a laser with the wavelength $\lambda_L=0.8\mu m$ is $\hbar\omega_L=1.55~\mathrm{eV}$. Laser photons scattered under the angle $\theta\ll 1/\gamma_e$ have the frequency (assuming linearly polarized laser pulse)
\begin{equation}
\omega\approx\frac{4\gamma_e^2}{1+\gamma_e^2\theta^2+a_0^2/2}\omega_L\mathrm{,} 
\label{photon_energy_eq}
\end{equation}
where $a_0=eA_L/mc^2$ is the normalized vector potential or laser pulse strength parameter (similar to the undulator strength parameter in the free-electron lasers). Here $A_L$ is the laser pulse vector potential amplitude in CGS units. Throughout the paper, we have neglected the recoil effect as the energy of the laser photon in the frame of reference of the electron ($\approx 2\gamma_e\omega_L)$ is still much smaller than the electron rest mass for the parameters of interest. \\ 
\\
\begin{figure}
   \centering
   \subfigure{\includegraphics[width=0.45\textwidth]{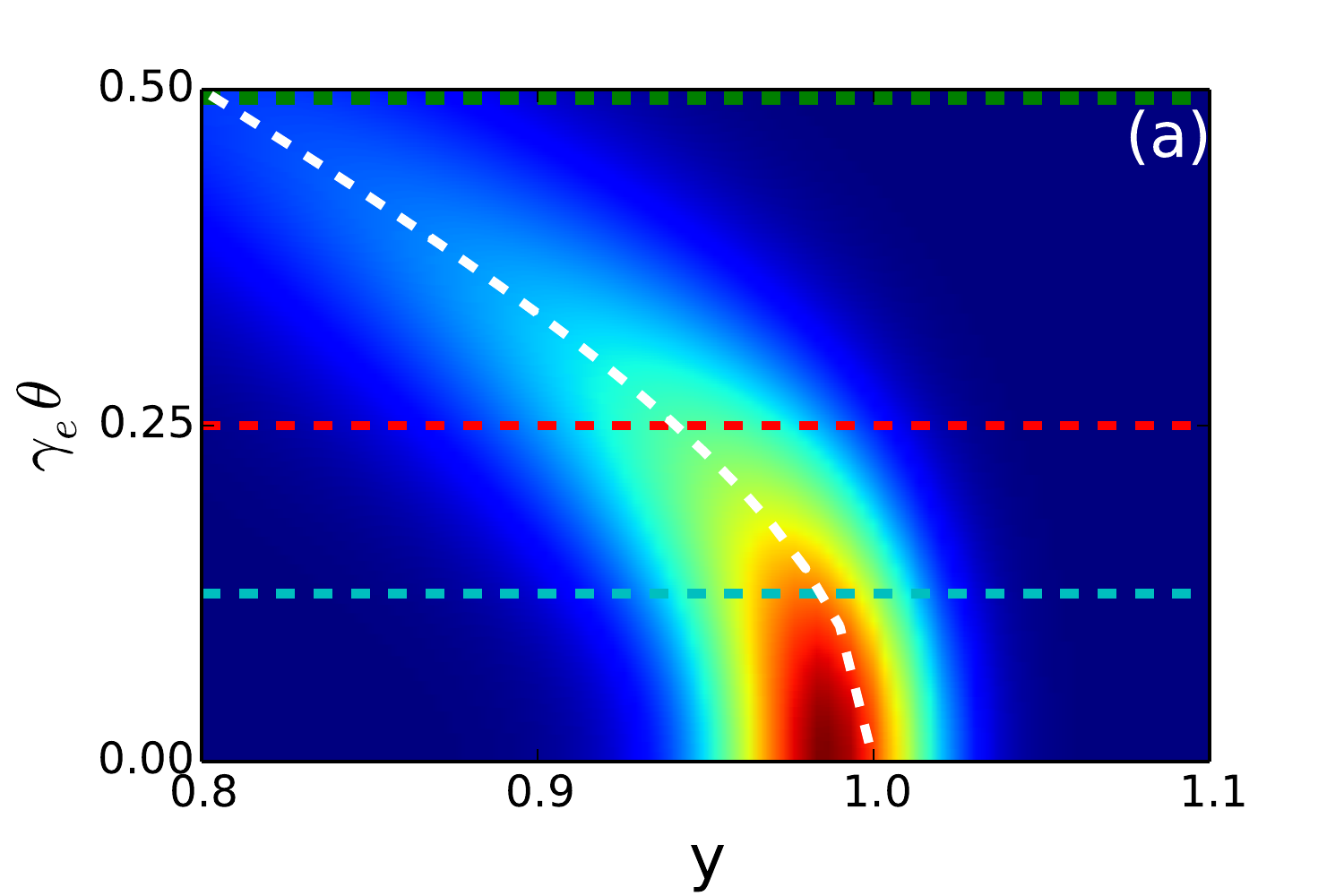}}
   \subfigure{\includegraphics[width=0.45\textwidth]{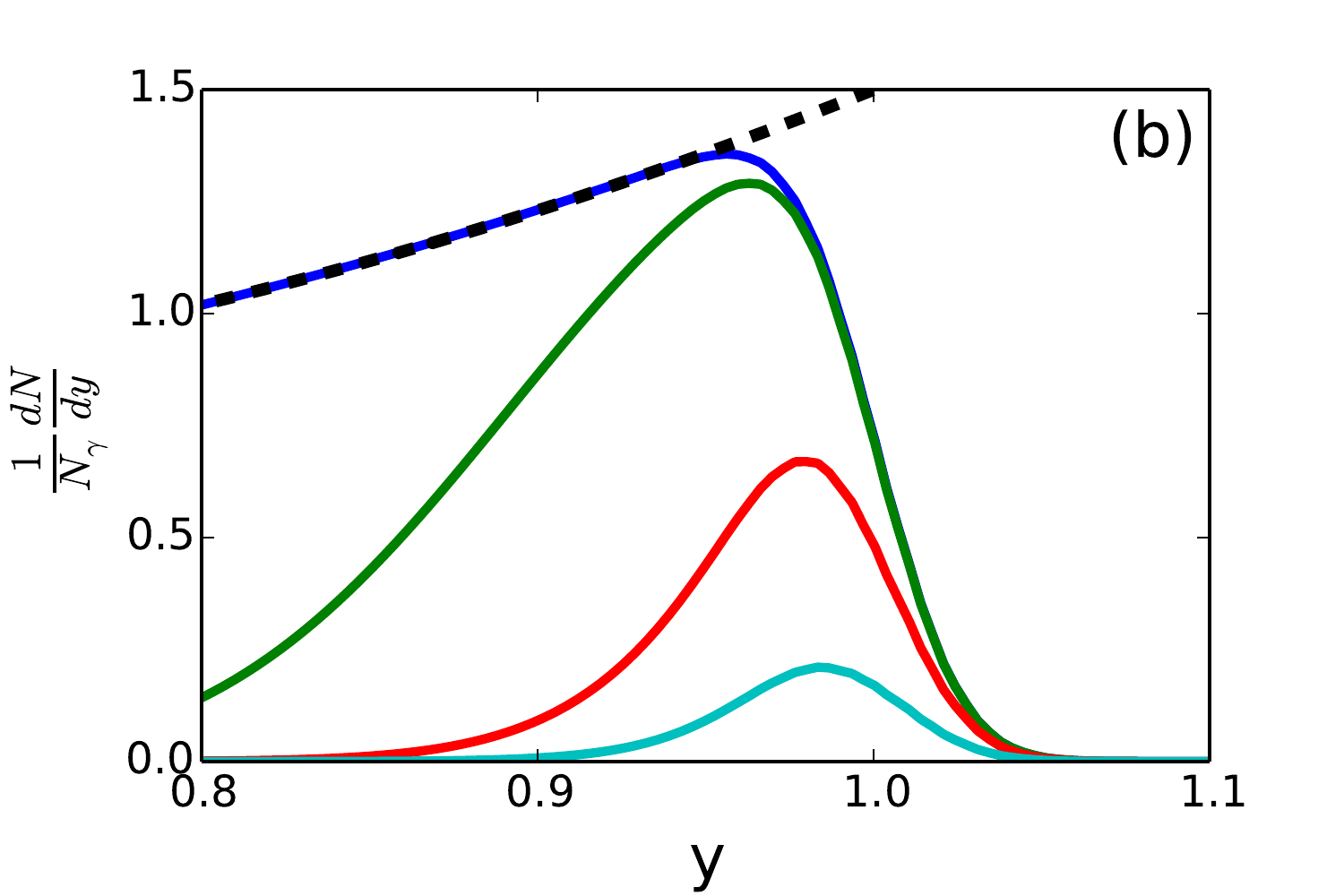}} 
     \caption{(a): Typical energy-angular spectrum of TS as a function of both the normalized energy $y=\frac{\omega}{4\gamma_e^2\omega_L}$ (horizontal axis) and inclination angle $\gamma_e\theta$ ($\theta=0$ is pointing in the direction opposite to the electron beam propagation axis). The white line shows the plot of normalized energy $y$ as a function of angle given by $y=\frac{1}{1+\gamma_e^2\theta^2}$ ($a_0$ is assumed small). Electron beam divergence and energy spread are taken into account and their influence is discussed in the text. (b): The photon energy spectra obtained with the help of collimation (integrating the spectrum from $0$ to $\theta_c$, the collimation angle) for four different cases: 1) no collimation (blue solid line); 2) $\gamma_e\theta_c=0.5$ (green solid line); 3) $\gamma_e\theta_c=0.25$ (red solid line) and 4) $\gamma_e\theta_c=0.125$ (cyan solid line). Corresponding collimation angles are also shown on (a) with dashed lines of same color.  }\label{typical_spec_figure}
\end{figure}
Typical energy-angular and photon energy spectra are presented in Fig.~\ref{typical_spec_figure}. For the plots of Fig.~\ref{typical_spec_figure}, electron beam divergence and energy spread were taken into account whereas the laser pulse was assumed to be infinitely long, non-divergent and of low intensity ($a_0\ll 1$). The source opening angle is roughly $1/\gamma_e$, with a bow-like energy-angular spectrum which is due to the $\theta$ term in eq.~(\ref{photon_energy_eq}). This leads to a broad integrated spectrum requiring collimation depending on the desired spectral width. Several examples of collimation and resulting photon energy distributions $\frac{1}{N_\gamma}\frac{dN}{dy}$ are presented on Fig.~\ref{typical_spec_figure}(b). Here $N_\gamma$ is the total number of generated photons so that the photon energy distribution is normalized to unity and 
\begin{equation}
y=\frac{\omega}{4\gamma_e^2\omega_L}=\frac{1}{1+\gamma_e^2\theta^2}
\label{y_eq}
\end{equation}
is the normalized photon energy.  To qualitatively assess the role of electron and laser beam parameters on TS spectrum one can look at equation (\ref{photon_energy_eq}). One can see that the frequency of the generated photon depends on four parameters: 1) the angle of propagation of the generated photon $\theta$; 2) the electron energy $\gamma_e$; 3) laser pulse amplitude $a_0$; and 4) laser frequency $\omega_L$. Realistic electron beams have a non-zero angular divergence. Electrons propagating under different angles will generate photon spectra peaked in the direction of their respective propagation. This broadens the integrated spectrum. Electron energy spread also leads to broadening as electrons in the beam having different energies will generate different photon energies in accordance with the $\gamma_e$ contribution in formula (\ref{photon_energy_eq}). The $a_0$ term in equation~(\ref{photon_energy_eq}) leads to additional hard photon beam broadening in the case when laser pulse has a non-constant intensity envelope. Indeed, according to the formula~(\ref{photon_energy_eq}) different frequencies will be generated at different times throughout the pulse (here we assume that there is no frequency chirp in the laser pulse). In experiments, depending on the desired bandwidth, one needs to keep $a_0$ as high as possible for maximizing the photon yield, but low enough to meet the bandwidth requirement. The requirements on the photon source hence put conditions on the laser and electron beams that can be used for generation.

\section{Spectral shape of the radiation}\label{spectral_shape_section}

In order to design a photon source one needs to know how parameters of
electron and laser beams influence the shape of the radiation
spectrum.  It is important to introduce the assumptions that are used
in calculation of the photon spectrum.  First of all, it is assumed
that the number of periods $N_{0}$ in the laser pulse is large,
$N_{0}\gg 1$.  For ideal electron beams with no energy spread and no
divergence, the normalized frequency width of the spectrum width is
given by $1/N_{0}$ (as in undulators~\cite{Schmueser2008}). For LPA
electron beams interacting with weak laser pulses $a_0\ll 1$ and photon sources of interest, however, the dominating
contributions to the spectral broadening are beam energy spread and
divergence. The contribution to the bandwidth for the case of a laser pulse with amplitude $a_0$ on the order of unity is discussed in Sec.~\ref{spectral_shape_intensity_section}. Thus, we first describe interaction of a single electron
with a long laser pulse (having delta-like frequency spectrum) and
then add electron beam divergence and energy spread into consideration
in Sections \ref{spectral_shape_div_section} and
\ref{spectral_shape_energy_spread_section} respectively.  These are
the dominant drivers of energy spread for Thomson photon sources.  The
approximations made by neglecting laser bandwidth and divergence are
small.  As we shall see in Sec.~\ref{yield_section}, the laser pulse
lengths required to generate strong scattering ($\sim$1
photon/electron) are picosecond scale such that the laser bandwidth is
less than or about 0.1$\%$, less than contributions due to energy
spread and divergence of LPA electron beams or, in many cases, even
conventional linacs.  Hence the dominating contributions to the
spectral broadening are energy spread and beam divergence.  Likewise,
the angular spread due to the focusing or diffraction of the laser
pulse is slight and can be neglected.  The frequency will be slightly
downshifted by the factor $\cos(\theta_d)$, where $\theta_d\approx
0.37\lambda_L/{w_{0,FWHM}}$ is diffraction angle for the
gaussian laser beam and $w_{0,FWHM}$ is the intensity full-width at
half maximum size of the laser beam.  In other words, spectral
broadening due to the laser beam diffraction is on the order of
${\theta_d^2}/{2}$.  Even for sharply focused laser beam with
$w_{0,FWHM}=4\lambda_L$, this contribution is on the order of $0.3\%$
and can be neglected for the most gamma sources of interest.  Under
given assumptions for the laser pulse one can separate the calculation
of the photon spectrum into two parts: 1) calculation of the
fractional number of photons within a given bandwidth taking into
account divergence and energy spread of electron beam, and 2)
calculation of the total number of generated photons taking into
account geometry of interaction, which is presented further in
Sec.~\ref{yield_section}.

Radiation from a particle moving with arbitrary trajectory can be
found using standard formula~\cite{Jackson1975}
\begin{equation}
\frac{d^2I}{d\omega d\Omega}=\frac{e^2\omega^2}{4\pi^2c}\left| 
\int\limits_{-\infty}^{+\infty}\mathbf{n}\times
\left(\mathbf{n}\times\boldsymbol\beta \right) 
e^{i\omega\left(t-\frac{\mathbf{nr}}{c} \right)}dt \right|^2\label{d2Idomegas_eq}\mathrm{,}
\end{equation}
where $d^{2} I$ is the energy radiated into the frequency band $d\omega$ 
and solid angle element $d\Omega$, 
$\mathbf{n}$ is a unity vector pointing from the electron
position to the detector, $\mathbf{r}$ is the electron radius-vector
and $\boldsymbol\beta=\mathbf{v}/c$ is the electron velocity.  
Detailed expressions for the energy radiated have been calculated for 
the case of a 
single relativistic electron colliding head-on with a linear 
polarized laser 
pulse with a flattop profile of amplitude $a_{0}$ consisting of $N_{0}$ perionds
\cite{Esarey1993, Leemans2005, Chen2013}.
In the limits
$a_0^2\ll 1$, $\gamma^2\gg 1$, and $\theta^2\ll 1$, the spectral energy
density of the radiation is given by
\begin{equation}
\frac{d^2I}{d\hbar\omega d\Omega}=\alpha\frac{\gamma_e^2 N_0^2a_0^2}
{\left(1+\gamma_e^2\theta^2 \right)^2}
\left[1-\frac{4\theta^2\gamma_e^2\cos^2\phi}
{\left(1+\gamma_e^2\theta^2 \right)^2} \right]
 R(\omega,\omega_R)\mathrm{,}\label{spec_general_eq}
\end{equation}
where $\alpha=e^2/\hbar c\simeq 1/137$ is a fine-structure constant,  $\theta$
and $\phi$ are polar and azimuthal angles respectively, and
\begin{equation}
\omega_R=\frac{4\gamma_e^2\omega_L}{1+\gamma_e^2\theta^2} 
\end{equation}
is the
resonance or peak frequency of the generated radiation.
Here, $R(\omega,
\omega_R)$ is the resonance function that depends on the exact pulse
shape.  For a flat-top laser pulse the resonance function is given by
\begin{equation}
R_F(\omega, \omega_R)=\left[\frac{\sin\left(\pi N_0
\left(\frac{\omega}{\omega_R}-1 \right) \right)}
{\pi N_0\left(\frac{\omega}{\omega_R}-1 \right)} \right]^2
\end{equation}
and the frequency width of the resonance function is $\Delta\omega =\int 
d\omega R_{F}=\omega_{R}/N_{0}$.

Of interest is the radiation collected by an axisymmetric detector 
placed along the axis some distance for the interaction point. In this 
case, Eq.\ (\ref{spec_general_eq}) can be averaged over the azimuthal 
angle $\phi$ giving
\begin{equation}
\frac{d^2I}{d\hbar\omega d\Omega}=\alpha\gamma_e^2  N_0^2 a_0^2
\left[
\frac{(1+\gamma_e^4\theta^4)}{(1+\gamma_e^2\theta^2 )^4}
\right]
 R(\omega,\omega_R)\mathrm{.}\label{spec_general_eq_ave}
\end{equation}

Since for $N_{0}\gg 1$, the radiation spectrum is narrowly peaked 
about the resonance frequency, the number of photons $N$ radiated per 
unit frequency and unit solid angle can be defined by 
$d^2 N/d\omega d\Omega =(1/\hbar\omega_{R})d^2 I/d\omega d\Omega$.
Integrating this expression over frequency and over $\phi$ gives
\begin{equation}
\frac{dN}{ dx}=\pi\alpha\gamma_e^2  N_0 a_0^2
\left[
\frac{(1+x^2)}{(1+x)^4}
\right]
\mathrm{,}\label{N_spec_general_eq_ave}
\end{equation}
where $x=\gamma^{2}\theta^{2}$.
From this equation, the total number of photons radiated in a 
$\theta= 1/\gamma$ cone is
\begin{equation}
N_{1} = (\pi/3) \alpha\gamma_e^2  N_0 a_0^2
\end{equation}
and the total number of photons radiated over all angles is twice 
this value, 
$N_{\gamma}= 2N_{1}$.


The above expressions can be generalized to other laser pulse 
profiles.
For the case of a Gaussian laser pulse with electric field
proportional to $a\propto \exp\left(-t^2/2\tau_L^2 \right)$, the
resonance function is given by
\begin{equation}
R_G(\omega,\omega_R)=2\pi\cdot e^{-(2\pi N_0)^2\cdot
\left(\frac{\omega}{\omega_R}-1\right)^2}\mathrm{,}
\end{equation}

Using eq.~(\ref{spec_general_eq}), one can obtain the following
expressions for the spectral energy and photon number per unit 
freqyency in the case of
the single electron (or an ideal beam of identical electrons)~\cite{Villa1996}:
\begin{eqnarray}
\frac{1}{4\gamma_e^2\hbar\omega_L N_\gamma}\frac{dI}{dy}
=\frac{3}{2}y\left( 1-2y(1-y)\right)\label{energy_spec_circ_eq}\\
\frac{1}{N_\gamma}\frac{dN}{dy}=\frac{3}{2}\left(1-2y(1-y) \right)
\label{photon_spec_circ_eq}\mathrm{,}
\end{eqnarray}
where 
$y=\omega_{R}/4\gamma_{e}^{2}\omega_{L}=1/(1+\gamma_{e}^{2}\theta^{2})$
is the normalized resonant frequency. 
Here $N_\gamma$ is the total radiated photon yield, which will be
calculated in Sec.~\ref{yield_section} for non-ideal beams.  A 
long laser pulse is assumed, $N_{0}\gg 1$, which means that a photon with a
specific energy has a specific angle of emission.  It
is clear that integrating the right part of
eq.~(\ref{photon_spec_circ_eq}) from $y=0$ to $y=1$ one gets unity.

\begin{figure}
   \centering
    \includegraphics[width=0.5\textwidth]{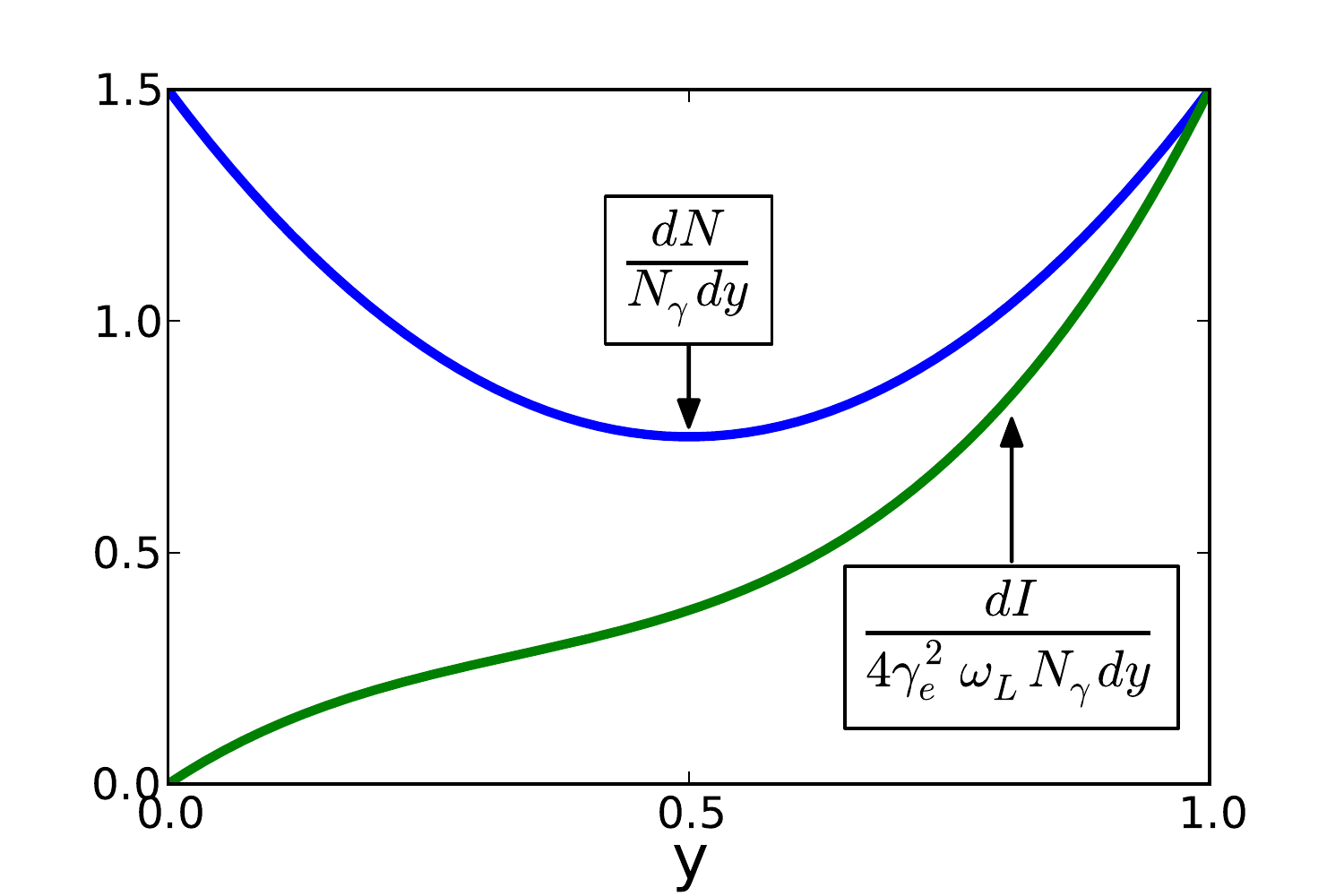}
     \caption{Photon spectrum (blue) and intensity spectrum (green) as
     functions of normalized photon energy $y$ obtained from
     eqns.~(\ref{photon_spec_circ_eq}) and (\ref{energy_spec_circ_eq})
     respectively.}\label{energy_spec_figure}
\end{figure}

Photon spectra given by eqns.~(\ref{energy_spec_circ_eq}) and
(\ref{photon_spec_circ_eq}) as functions of $y$ are plotted in
Fig.~\ref{energy_spec_figure} with green and blue lines respectively.
One can see that though most of the generated energy is concentrated
near the maximum frequency ($y=1$), only a small fraction of all
generated photons have energy close to the maximum photon energy of
$E_{max}=4\gamma_e^2\hbar\omega_L$.  In fact, one can calculate the
relative number of photons in a given bandwidth near the maximum
frequency exactly.  Denoting $\kappa$ as the relative FWHM (full width
at half maximum) bandwidth of the hard photon source near the maximum
frequency, and integrating eq.~(\ref{photon_spec_circ_eq}) from
$y=1-\kappa$ to $y=1$ one obtains the following expression for the
number of photons in the bandwidth $\kappa$
\begin{equation}
N_\kappa=N_\gamma \kappa \left(\kappa^2-3/2\kappa+3/2 \right)\mathrm{.}
\end{equation}
For example, taking a required bandwidth $\kappa=0.02$ or, in other
words $2\%$, one can calculate that approximately 3$\%$ of all
generated photons lie in this bandwidth and thus the majority of
photons have energies outside of the required bandwidth.  A bandwidth
of 10$\%$ contains 14$\%$, or nearly 5 times more photons than in the
$2\%$ bandwidth case, illustrating source tradeoffs.  It is important
to emphasize that the considered case is ideal: electron beam does not
have energy or angular spread, laser pulse bandwidth is infinitely
narrow, and is a plane wave.  Nevertheless, the case considered in
this section provides important estimates for the TS photon source.


\subsection{Electron beam divergence effects}\label{spectral_shape_div_section}
Angular spread in the electron beam distribution will lead to bandwidth broadening as particles moving under different angles will each generate maximum frequency of $\omega_{max}=4\gamma_e^2$ in the direction of their propagation and not necessarily along the $z$ axis. It is important to take electron beam divergence into account as it can often be the dominant contribution to energy spread for conventional linac sources \cite{Albert12, Leemans1996}. Specific needs for LPA sources are presented in Sec.~\ref{using_LPA_section}. Without losing generality we consider the case of a circularly polarized laser pulse interacting with an electron beam with some angular spread. Polarization of the laser pulse will also influence the polarization of the generated X-ray photons, but the total number of generated photons will remain the same.
Consider a round electron beam with angular distribution function given by
\begin{equation}
f_e\left(\theta_e, \phi \right)=\frac{1}{2\pi \sigma_\theta^2} \exp\left({-\frac{\theta_e^2}{2\sigma_\theta^2}}\right)\mathrm{,}
\end{equation}
where $\gamma_e\sigma_\theta\ll 1$ is the RMS electron beam divergence in any transverse plane. The electron energy spread is considered to be zero. The angle and energy spectrum of the generated $\gamma$ source can be then found by integration of eq.~({\ref{photon_spec_circ_eq}}) with respect to the electron propagation angles and reads~\cite{Villa1996}
\begin{equation}
\frac{1}{N_\gamma}\frac{d^2N}{dy \theta d\theta}=\frac{3}{2\sigma_\theta^2}(1-2y(1-y))\exp\left(-\frac{\theta^2+\tilde\theta^2}{2\sigma_\theta^2} \right)\mathrm{I}_0\left(\frac{\theta\tilde\theta}{\sigma_\theta^2}\right)\mathrm{,}\label{angle_energy_spec_circ_eq}
\end{equation}
where I$_0$ is the modified Bessel function and $\tilde\theta^2=\frac{1-y}{\gamma_e^2y}$. One can see that the width of the spectrum is governed by the electron beam divergence $\sigma_\theta$. It is convenient and common to use the FWHM bandwidth instead of RMS bandwidth. Using eq.~(\ref{angle_energy_spec_circ_eq}) and denoting the desired relative FWHM bandwidth of the $\gamma$ source as $\kappa$, one can derive the following approximate condition for the electron beam FWHM divergence
\begin{equation}
\gamma_e\sigma_{\theta\mathrm{,FWHM}}<2\sqrt{\kappa}\mathrm{.}
\label{sigma_theta_condition_eq}
\end{equation}

\subsection{Influence of electron beam energy spread}\label{spectral_shape_energy_spread_section}

Energy spread of the electrons in the beam leads to photon source bandwidth broadening because electrons with different energies $\gamma_e$ will generate different photon energies. The effect of the electron energy spread only (assuming an electron beam with no divergence) on the on-axis hard photon source spectrum bandwidth can be easily estimated. Again, denoting $\kappa$  the desired relative FWHM bandwidth and differentiating eq.~(\ref{photon_energy_eq}) with respect to $\gamma_e$, one gets the following condition for the electron beam FWHM energy spread, assuming for simplicity a Gaussian distribution for $\gamma_e$:
\begin{equation}
\frac{\sigma_{\gamma_e\mathrm{,FWHM}}}{\gamma_e}<\frac{\kappa}{2}\mathrm{,}
\label{sigma_gamma_condition_eq}
\end{equation}
so that an electron beam with 5$\%$ FWHM energy spread will generate hard photon source with at least 10$\%$ FWHM bandwidth. Typically, LPA electron beams have energy spread on the one percent level \cite{Esarey2009,Faure2004,Geddes2004,Mangles2004,Leemans2006} and are thus usable for generating few percent narrow bandwidth photon sources.
Combining equation (\ref{sigma_gamma_condition_eq}) with the condition for the $\sigma_{\theta\mathrm{,FWHM}}$ from eq.~(\ref{sigma_theta_condition_eq}) one can get the combined approximate condition for both energy and angular electron beam spreads
\begin{equation}
\sqrt{\frac{\gamma_e^4\sigma_{\theta\mathrm{,FWHM}}^4}{16}+\frac{4\sigma_{\gamma_e\mathrm{,FWHM}}^2}{\gamma_e^2}}<\kappa\mathrm{.}
\label{sigma_theta_sigma_gamma_condition_eq}
\end{equation}

Unfortunately, to formally include the effects of both energy spread and the electron beam divergence as it was done in the previous section one at present has to use numerical integration.  An example of numerical integration of eq.~(\ref{angle_energy_spec_circ_eq}) taking into account both the electron beam divergence ($\gamma_e\sigma_{\theta\mathrm{,FWHM}}\approx0.24$) and electron beam energy spread ($\sigma_{\gamma\mathrm{,FWHM}}/\gamma_e\approx0.024$) is presented in Fig.~\ref{typical_spec_figure}. In this case the on-axis bandwidth (cyan line) was approximately 5 percent due to the broadening from both the electron beam divergence and electron beam energy spread, in good agreement with with the approximate solution of eq.~(\ref{sigma_theta_sigma_gamma_condition_eq}), which gives 5$\%$ for the parameters described above. 

\subsection{Estimation of the collimation angle and relative photon number}\label{estimation_collimation_section}
\begin{figure}
   \centering
   \subfigure{\includegraphics[width=0.45\textwidth]{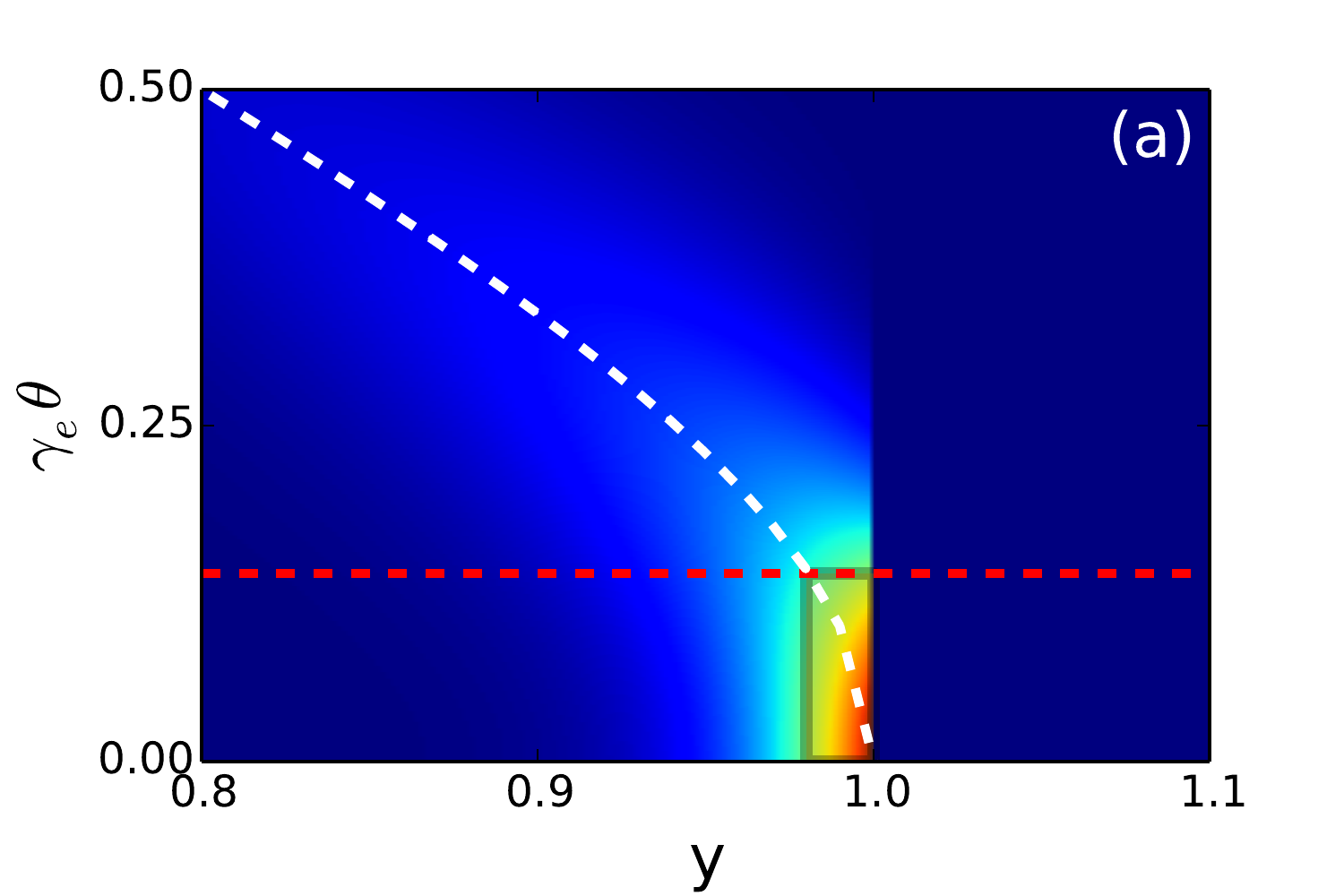}}
   \subfigure{\includegraphics[width=0.45\textwidth]{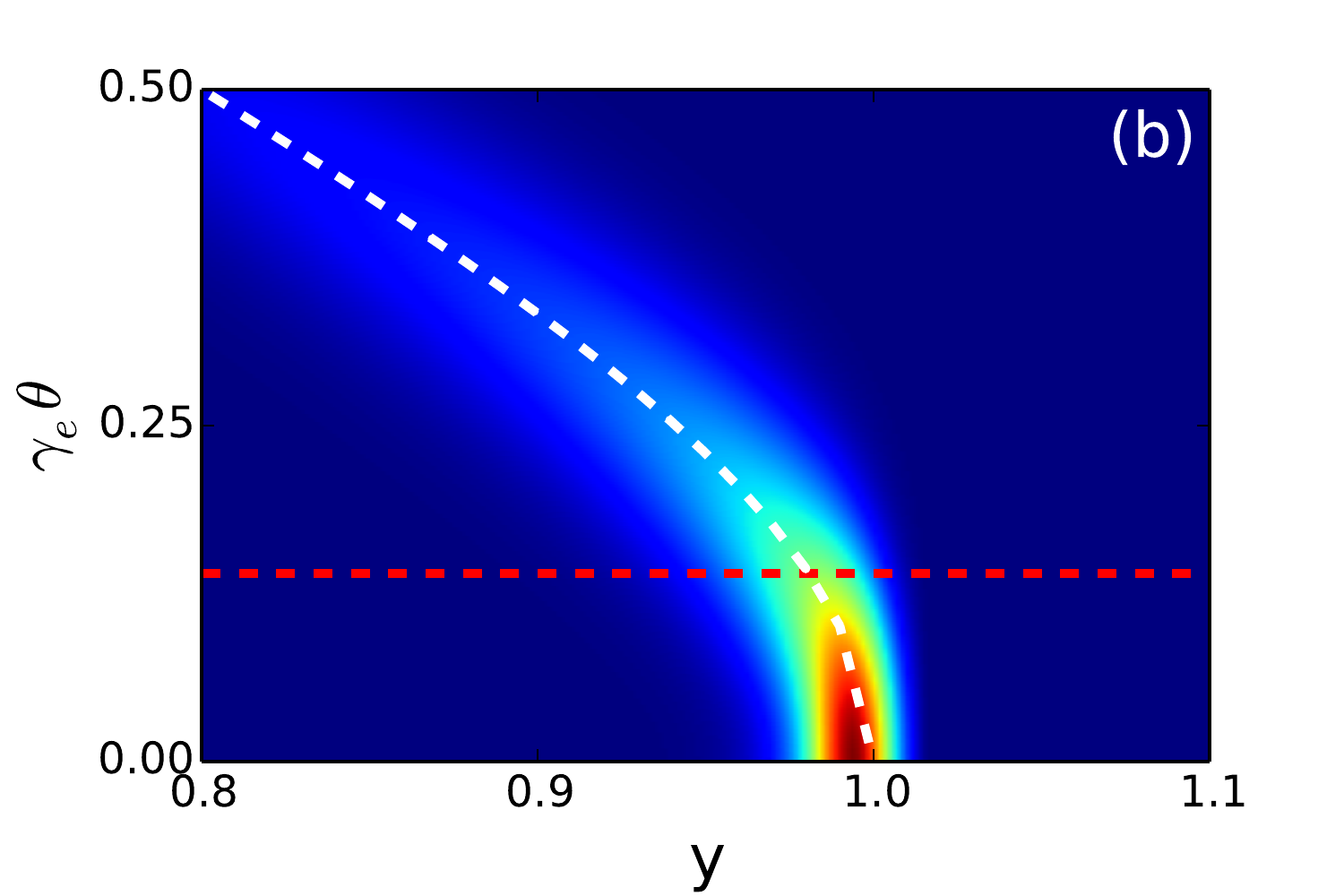}} 
     \caption{Energy-angular spectra of TS as functions of both the normalized energy $y=\frac{\omega}{4\gamma_e^2\omega_L}$~(horizontal axis) and inclination angle $\gamma_e\theta$ for two cases: (a) angular divergence dominated electron beam (electron energy spread is negligibly small) and (b) energy spread dominated electron beam (electron beam angular divergence is negligibly small). Photon source bandwidth $\kappa=0.02$ is the same for both plots and eq.~(\ref{sigma_theta_sigma_gamma_condition_eq}) is satisfied. The white line shows the plot of single electron normalized energy $y$ as a function of angle given by $y=\frac{1}{1+\gamma_e^2\theta^2}$ ($a_0$ is assumed small) and serves to guide the eye. In (a) the green colored area outlines the approximate area of integration for photon number estimation described in text.}\label{gamma_theta_comparison_figure}
\end{figure}

The energy-angle correlation of the Thomson spectrum means that collimation is required to achieve narrow bandwidth even for ideal electron and laser beams. To design a source to produce a given bandwidth $\kappa$ one needs to estimate: 1) the collimation angle $\theta_c$ and 2) the relative number of photons lying in this bandwidth. For the case of angular divergence dominated electron beams, such that the contribution of the angular spread is much higher than the contribution of the electron energy spread, the photon spectrum can be obtained with the help of eq.~(\ref{angle_energy_spec_circ_eq}). For the case of $\kappa=0.02$ the spectrum is presented in Fig.~\ref{gamma_theta_comparison_figure}~(a). One can estimate the collimation angle in the following way. Generated photon frequency is given by eq.~(\ref{photon_energy_eq}). We can approximate the collimation angle $\theta_c$ as such angle for which the photon frequency (for the case of $a_0\ll 1$) is equal to $\omega=\left(1-\kappa\right)\cdot 4\gamma_e^2\omega_L$. In other words,
\begin{equation}
\frac{1}{1+\gamma_e^2\theta_c^2}=1-\kappa\mathrm{.}
\end{equation}
Using eq.~(\ref{sigma_theta_condition_eq}) one can write the following expression for the collimation angle
\begin{equation}
\gamma_e\theta_c=\frac{\gamma_e\sigma_{\theta,FWHM}}{2}=\sqrt{\kappa}\mathrm{.}\label{collimation_angle_eq}
\end{equation}
For $y=1$, which is the maximum generated energy ($\omega=4\gamma_e^2\omega_L$), the angular dependence of the spectrum obtained from eq.~(\ref{angle_energy_spec_circ_eq}) is given by
\begin{equation}
\frac{1}{N_\gamma}\frac{d^2N}{dy \theta d\theta}|_{y=1}\propto \exp\left(-\frac{\theta^2}{2\sigma_\theta^2} \right)\mathrm{.}\label{yeq1spec_eq}
\end{equation}
One can easily check that the collimation angle corresponds to the angle such that the exponential term in eq.~(\ref{yeq1spec_eq}) is equal to 1/2. The corresponding collimation angle is outlined with red dashed line of Fig.~\ref{gamma_theta_comparison_figure}. 

\begin{figure}
   \centering
    \includegraphics[width=0.5\textwidth]{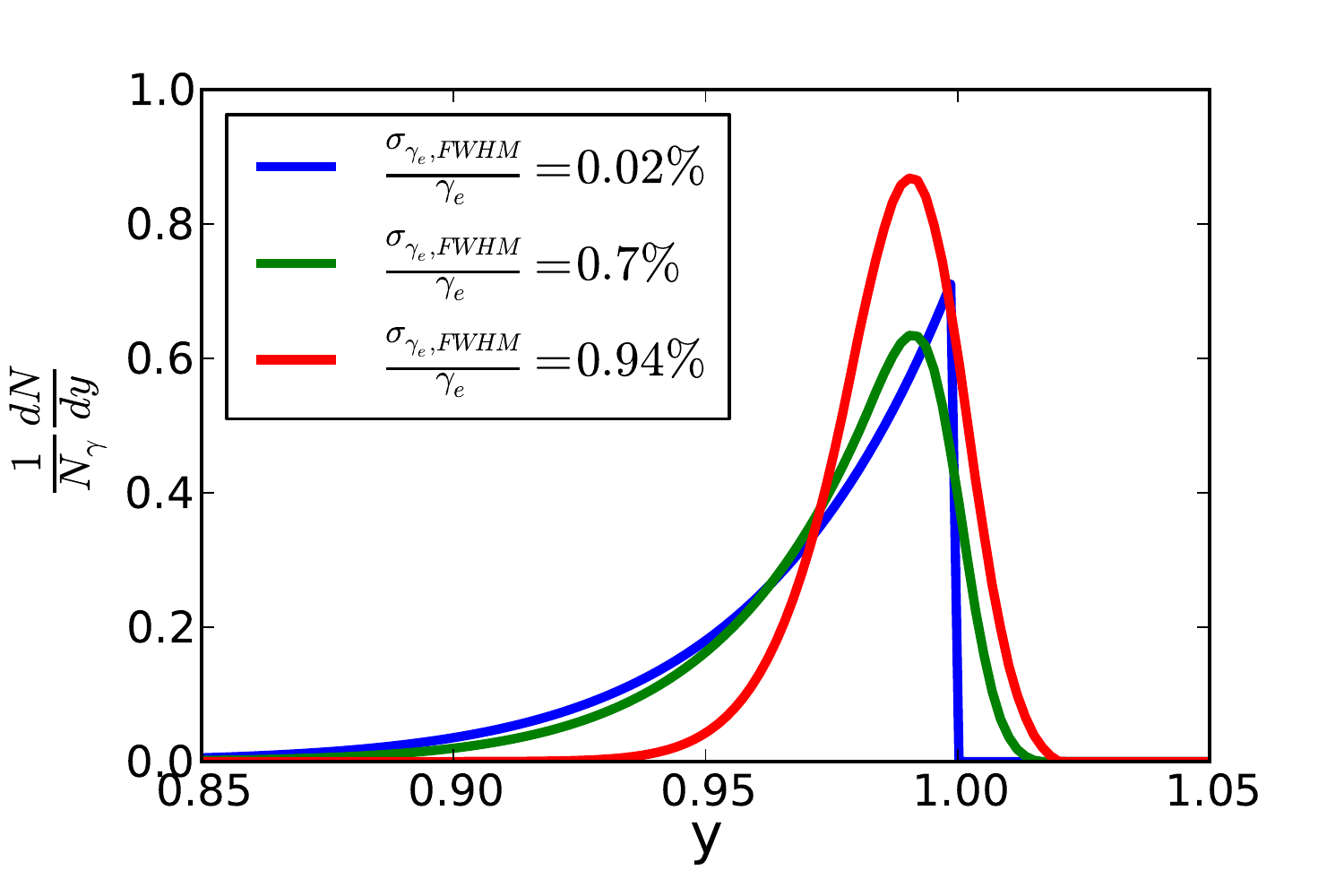}
     \caption{Photon spectra as function of normalized photon energy $y$ for different values of electron energy spread. Electron beam angular spread is changed according to eq.~(\ref{sigma_theta_sigma_gamma_condition_eq}) in such a way that the total source bandwidth is 2 percent.}\label{spectra_2_percent_compare_fig}
\end{figure}

One can now estimate approximately the relative number of photons lying in the bandwidth $\kappa$ by multiplying the peak value of the spectrum $\left(\frac{3}{2\sigma_\theta^2}\right)$ by bandwidth $\kappa$ and angular area $\frac{1}{2}\theta_c^2$ (corresponding area is outlined on Fig.~\ref{gamma_theta_comparison_figure}~(a) with the green rectangle). Using eq.~(\ref{collimation_angle_eq}) one can get the following estimate
\begin{equation}
\frac{N_\kappa}{N_\gamma}\approx  \frac{3}{2\sigma_\theta^2}\cdot\kappa\cdot\frac{1}{2}\theta_c^2\approx \kappa\mathrm{.}\label{rough_estimate_photon_eq}
\end{equation}
For example, in 2 percent bandwidth there are roughly 2 percent of all generated photons. We have found empirically with the help of numerical simulations that this estimate works well for both angular divergence dominated and energy spread dominated electron beams. The photon spectrum for the case of the energy spread dominated electron beams is presented in Fig.~\ref{gamma_theta_comparison_figure}~(b) and demonstrates the difference in the shape of the spectrum. Figure \ref{spectra_2_percent_compare_fig} shows the integrated (from 0 to collimation angle) photon energy spectra for different values of electron energy spread such that the total broadening is 2 percent and condition of eq.~(\ref{sigma_theta_sigma_gamma_condition_eq}) is satisfied. In all cases the relative number of photons was approximately 2.5 percent in good agreement with the estimate of eq.~(\ref{rough_estimate_photon_eq}). These rough estimates for the collimation angle and relative photon number can be used in source design. For more accurate answers, numerical integration can be applied.

\subsection{Influence of laser pulse intensity and multiple scattering}\label{spectral_shape_intensity_section}
Generally, one wants to maximize the laser photon flux by increasing the laser pulse amplitude $a_0$ to maximize the scattered photon yield. However, for higher values of $a_0$ there are several effects that decrease the quality of the source. First, the spectrum is broadened due to the appearance of sub-structures in the spectrum~\cite{Brau2004,Seipt2011,Heinzl2010, Maroli2013, Hartemann1996}. For the case of $a_0>1$ laser photons are wasted for generation of harmonics and scattering at undesired energies which leads to inefficient usage of laser energy. Hence, one needs to find a proper balance between the nonlinearity that one can tolerate and the total photon yield.

\begin{figure}
   \centering
    \includegraphics[width=0.45\textwidth]{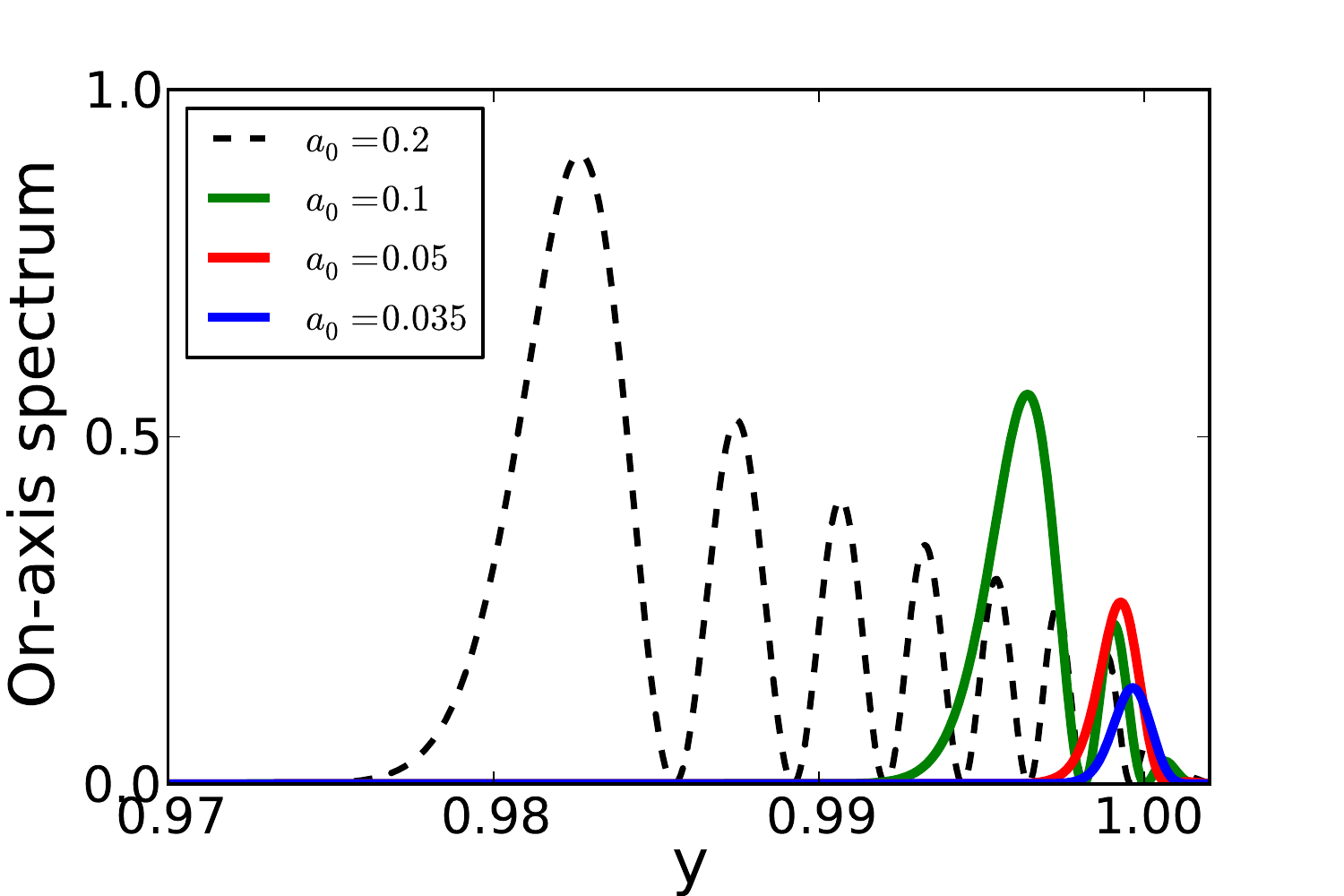}
     \caption{Normalized on-axis spectrum as a function of normalized photon energy $y$ for different laser pulse amplitudes: 1) $a_0=0.2$ (black dashed line); 2) $a_0=0.1$ (green solid line); 3) $a_0=0.05$ (red solid line); 4) $a_0=0.035$ (blue solid line).}\label{spec_ampl_figure}
\end{figure}

Results in the previous sections were obtained assuming that the laser pulse amplitude $a_0$ is much smaller than unity so that the scattering is linear. In the non-linear case eqns.~(\ref{energy_spec_circ_eq},\ref{photon_spec_circ_eq}) are not valid and one has to calculate the spectrum using eq.~(\ref{d2Idomegas_eq}). The frequency of the backscattered light is approximately given by eq.~(\ref{photon_energy_eq}), where $a_0$ can also be a function of time depending on the laser pulse envelope. This leads to photon source bandwidth broadening and substructures in the spectrum \cite{Brau2004,Ghebregziabher2013, Hartemann1996} as well as appearance of harmonics. Example spectra calculated from eq.~(\ref{d2Idomegas_eq}) using the numerical code VDSR \cite{Chen2013} for a single electron interacting with a gaussian plane wave with FWHM duration of 800~fs are presented in Fig.~\ref{spec_ampl_figure} for different laser pulse amplitudes: 1) $a_0=0.035$ (blue solid line); 2) $a_0=0.05$ (red solid line); 3) $a_0=0.1$ (green solid line); 4) $a_0=0.2$ (black dashed line). Increasing the laser pulse amplitude from $a_0=0.035$ to $a_0=0.05$ (thus doubling the number of laser photons) has not changed the spectrum shape much and the amplitude of the spectrum therefore increased by a factor of 2 proportionally to the number of laser photons. Further increase of laser pulse amplitude to $a_0=0.1$ (thus quadrupling the number of laser photons compared to the case of $a_0=0.05$) leads to appearance of the sidebands and broadening. The amplitude of the spectrum increased only by about a factor of 2 although the laser photon number increased by a factor of 4. Further increase of laser pulse amplitude to $a_0=0.2$ leads to considerable sidebands for a single electron, and in general this broadening is on the order of $a_0^2/2$.  For scattering from an electron beam this term should be kept small relative to the electron energy spread and divergence effects to minimize bandwidth, e.g. the broadening at $a_0=0.2$ induces 2$\%$ energy spread comparable to that from 1$\%$ electron energy spread. It is important to mention that appropriate laser pulse chirping may lead to narrowing of the bandwidth as discussed based on a single set of numerical parameters in\cite{Ghebregziabher2013}. Our analytical derivations and simulations of bandwidth reduction using laser chirp will be presented in a separate publication (see also Ref.~\cite{Terzic2014} for a discussion of chirp for bandwidth control). Another possible way to mitigate the broadening due to the laser pulse amplitude is using the flat-top laser pulses having almost rectangular shape in every direction~\cite{Tomassini2005}. In this case one can push the laser pulse amplitude to $a_0\approx 1$ and use the standard theory of undulators~\cite{Schmueser2008, Huang2007,Esarey1993}.

Throughout the paper we have neglected the recoil effect on the electron as it emits a single hard photon (with "average" photon energy of $2\gamma_e^2$) as the energy loss is small compared to electron energy
\begin{equation}
\frac{\Delta\gamma_e}{\gamma_e}\approx 2\gamma_e\frac{\hbar\omega_L}{m_e c^2}\ll 1\mathrm{.}\label{energy_loss_multiple_scat_eq}
\end{equation}
However, if the electron emits multiple photons during the interaction, the cumulative effect of recoil on the spectrum should be taken into account. Denoting $N_{sc}$ as the number of times an electron scattered a photon during the interaction, the product of $N_{sc}$ and the energy loss given by eq.~(\ref{energy_loss_multiple_scat_eq}) should be less than desired source bandwidth. The value for $N_{sc}$ can be approximated using the cross-section formalism and will be presented in Sec.~\ref{yield_section}. As an example, consider TS of 0.8 micrometer laser light from electrons with $\gamma_e=500$ leading to photon energy of roughly 1.5~MeV. In this case, an average of 8 scatterings would lead to photon energy change on the order of 2 percent.

Given the required FWHM bandwidth $\kappa$ one can now write the following approximate condition taking into account electron beam divergence and energy spread as well as laser pulse amplitude and multiple scattering, and adding them in quadratures
\begin{equation}
\sqrt{\frac{\gamma_e^4\sigma_{\theta\mathrm{,FWHM}}^4}{16}+\frac{4\sigma_{\gamma_e\mathrm{,FWHM}}^2}{\gamma_e^2}+\frac{a_0^4}{4}+\left[N_{sc}\cdot 2\gamma_e\frac{\hbar\omega_L}{m_e c^2} \right]^2}<\kappa\mathrm{.}
\label{requirements_bandwidth_eq}
\end{equation}
As discussed in Sec.~\ref{estimation_collimation_section}, the collimation angle is approximately given by $\gamma_e\theta_c\approx\sqrt{\kappa}$ and a fraction of photons of approximately $\kappa$ is lying in this bandwidth. These results provide important estimates and constraints on the electron beam for designing a photon source with specified bandwidth $\kappa$ and were also verified with the help of numerical simulations.

\section{LPA electron beams for narrow-bandwidth photon sources}\label{using_LPA_section}
Several experiments using TS from LPA electron beams demonstrated the generation of broad bandwidth (more than 20 percent) X- and $\gamma$-ray photon sources~\cite{Schwoerer2006, TaPhuoc2012,Chen2013a}. Some applications, such as NRF and photo fission for active nuclear interrogation of cargo and nuclear waste or nuclear physics studies benefit from narrow bandwidth of the photon source. In this section we address the ways to control LPA electron beam divergence for obtaining the narrow-bandwidth photon sources and a compact beam dump, needed to dispose of particles after the interaction.

\subsection{Control of the electron beam divergence}
Laser plasma accelerators produce low emittance, GeV-level electron beams~\cite{Leemans2006, Plateau2012, Weingartner2012} that allow generation of multi-MeV photons suitable for NRF and photo fission experiments.  Recently measured normalized electron beam emittance is on the order of $\varepsilon_{e,n}\approx 0.1\mathrm{~mm}\cdot\mathrm{mrad}$ for electrons with $\gamma_e\approx 1000$~\cite{Plateau2012}. Electron beam transverse size inside an LPA is on the order of $\sigma_{e,0}\approx 0.1\mu$m and divergence is on the level of $\sigma_{e,\theta}\approx 1$~mrad because of the strong focusing inside the plasma wave. As a result their divergence is rather large for generation of percent-level narrow-bandwidth photon sources. In this case $\gamma_e\sigma_{\theta}\approx 1$ and according to the bandwidth condition given by eq.~(\ref{requirements_bandwidth_eq}) the minimum photon source bandwidth is on the order of 20$\%$. For generation of 2 percent narrow photon sources for NRF studies, the divergence has to be decreased by approximately an order of magnitude. There are several possible methods for decreasing the electron beam divergence.\\
1) \textit{Quadrupole magnet lenses.} Perhaps the most straightforward and commonly used method is refocusing the electron beam using quadrupole magnet lenses. In this technique, the electron beam size is blown up keeping the beam emittance constant, thus reducing the divergence. This method has been demonstrated to successfully function and is fairly easy to implement. According to the results presented in \cite{Osterhoff2010, Weingartner2011}, the setup size for miniature permanent quadrupole magnets (PMQ) is on the order of couple of meters. The same setup is needed for a proposed table-top FEL~\cite{Gruner2007,Maier2012,Huang2012,Schroeder2006}. 
\\
2) \textit{Controlled injection.}
Currently, a large amount of research in the area of LPA is dedicated to investigation and development of novel electron injection methods. Control of electron bunch position inside the accelerating structure (plasma wave) can lead to reduction in both energy spread and divergence of the electron beam as demonstrated in experiments and theoretical works~(see \cite{Esarey2009} and references therein). Promising methods currently being under consideration include so-called ionization injection, where addition of high-Z gas species and additional laser pulses leads to decrease in the electron beam divergence~\cite{Moore1999,Yu2014,Hidding2012,Chen2006,McGuffey2010} and use of colliding pulses or density ramps. Research is required to answer the question of whether the electron beam emittance can be reduced at least 10 times compared to the present day values. Similar reduced emittance is also needed for the High Energy Physics applications such as electron-positron colliders discussed in \cite{Leemans2009, Schroeder2010}.\\
3) \textit{Downramp.} It is possible to reduce the electron beam divergence by reducing the focusing forces of the plasma wave and thus increasing the beam size adiabatically~\cite{Sears2010a}. Assuming that the plasma focusing force is linear and thus the normalized emittance is conserved, increase in the beam size leads to the decrease of the beam divergence. This can be achieved by controlled density profile with negative gradient - downramp. If the electron beam is matched inside the plasma structure, then its radius and divergence in the blowout regime are given by~\cite{Esarey2002}
\begin{eqnarray}
r_{m}=\sqrt{\frac{\varepsilon_n\lambda_p}{\pi\sqrt{2\gamma_e}}}\label{matched_radius_eq}\\
\gamma_e\sigma_\theta=\sqrt{\pi\sqrt{2\gamma_e}}\sqrt{\frac{\varepsilon_n}{\lambda_p}}\mathrm{,}\label{div_eq}
\end{eqnarray}
where $\varepsilon_n$ is the normalized beam emittance and $\lambda_p$ is the plasma wavelength. As $\lambda_p\propto n_e^{-1/2}$, where $n_e$ is the plasma density,  in order to increase the electron beam size $m$-fold, plasma density has to be reduced $m^4$ times. LPAs producing 0.5~GeV electron beams with 0.1~mm$\cdot$mrad normalized emittance operate using $n_e=5\cdot 10^{18}\mathrm{~cm}^{-3}$ with plasma wavelength equal to $\lambda_p\approx 15\mu$m. In order to decrease the electron beam divergence 10 times, the density must adiabatically (i.e. the local scale length must allow at least one betatron oscillation) drop to approximately $n_e\approx 5\cdot 10^{14}\mathrm{~cm}^{-3}$. This leads to a length of the plasma downramp on the order of 1 meter and might be difficult to achieve in experiments.\\
4) \textit{Plasma lens.} Plasma lenses based on axisymmetric electrostatic forces, generated by expelling all or part of electrons from the plasma region may be used for focusing of the electron beams~\cite{Panofsky1950, Chen1987, Goncharov2013}. In the case of complete electron blow-out the radial electric field in a plasma channel (lens) is given by
\begin{equation}
E_r=\frac{m_e c^2 k_p^2}{e}\cdot\frac{r}{2}\mathrm{,}\label{radial_field_eq}
\end{equation}
where $k_p=\sqrt{4\pi r_e n_e}$ is the plasma wavenumber with $r_e$ being the classical electron radius and $n_e$ the density of the plasma. Using eq.~(\ref{radial_field_eq}) one can roughly estimate the parameters of the plasma lens (in the thin-lens approximation)
\begin{equation}
n_e\cdot l=2.84\cdot 10^{11}\left[\mathrm{cm}^{-1} \right]\cdot\frac{\gamma_e}{d}\mathrm{,}
\end{equation}
where $l$ is the length of the plasma lens and $d$ is the length of the drift space between the electron source (LPA) and the lens. For $\gamma_e=1000$, 1~mrad electron beam divergence and a drift space of 1mm, in order to collimate the electron beam a plasma lens with parameters $n_e\cdot l=2.84\cdot 10^{15}$cm$^{-2}$ is required. For example, a 300~$\mu m$ slab of plasma with density $10^{17}$cm$^{-3}$ would suffice. It is important to mention that plasma lens can be created by the same laser that drives the LPA in the case when drift space $d$ is smaller than the Rayleigh length of the laser pulse as in the provided example. This can make experimental setup significantly more compact. A conceptual experimental setup is presented on Fig.~\ref{tentative_setup_figure}.

Other methods, such as, for instance, radiative beam cooling~\cite{Telnov1997,Esarey2000,Yokoya2000} or controlling focusing forces using different laser beam modes~\cite{Cormier-Michel2011} may also be used. The most straightforward approach is use of PMQs, whereas higher performance may be possible using the plasma lens. More research in this area is needed.

\subsection{Compact beam dump}

Disposal of the high energy particle beam after photon production imposes the use of large and heavy "beam dumps" that usually prevent portability of the photon source, limiting applicability. For portability, constraints of size and weight require acceleration of the electron beam in a short distance and also disposal of its energy (after photon production) in a way that minimizes size and the use of heavy materials. High beam currents are also required to meet application needs for photon flux, compounding the problem. As mentioned above, LPAs have been demonstrated to produce high-quality electron beams at the required energies in cm-scale distances, fulfilling the need for compact acceleration. With such small accelerators, the size of a photon source would be dominated by conventional methods for disposal of the electron beam which require heavy shielding. For the beam energies required to produce MeV photons, this shielding is of room size and can preclude transportable operation.  In principle, the same structure used for acceleration in an LPA can be used to decelerate the electron beam by appropriately phasing the beam in the plasma wake.  This can theoretically decelerate the beam over the same cm-scale distance as required for acceleration.  Limitations include dephasing with regard to deceleration and focusing as well as energy spread from non-uniform deceleration.

 In laser-driven plasma accelerators, a laser displaces electrons in a plasma channel, initiating plasma oscillations and resulting in a succession of positively and negatively electrically charged regions behind the laser (or ``wake"). 
 The alternating polarity within the wake generates very strong (typically GV/m) longitudinal and transverse electric fields of alternating sign. 
 An electron beam located at the appropriate phase behind the laser will be both focused transversely and accelerated to (or decelerated from) high energies over a very short distance. 
The accelerating and focusing fields are driven by the ponderomotive force of the laser pulse $F\simeq-m_e c^2\nabla a^2/2$.
The longitudinal field is of the order of $E_p$(V/m)$\simeq96\sqrt{n_0(\mathrm{cm^{-3}})}$ with $n_0$ the plasma electron density, which can reach several orders magnitude higher amplitudes than with conventional acceleration techniques.

\begin{figure}
\begin{centering}
\includegraphics[scale=0.3]{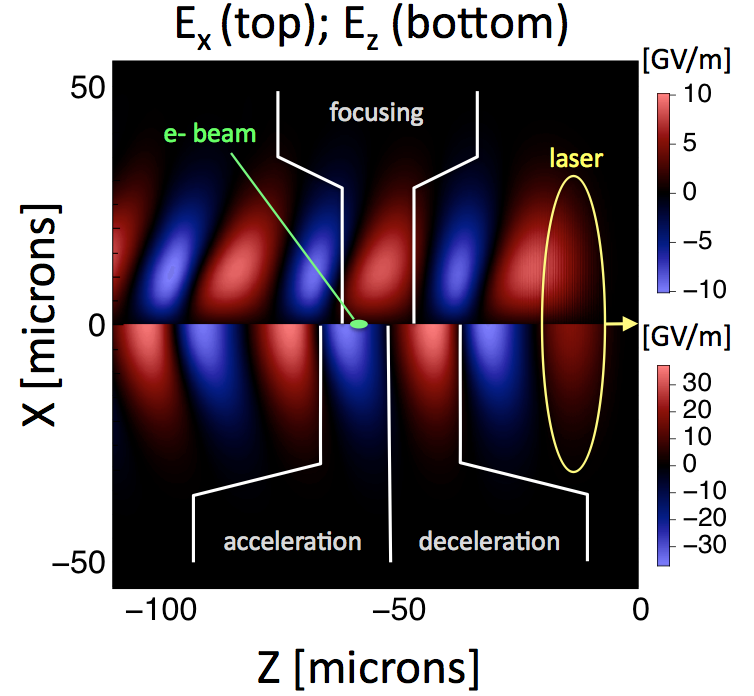}
\par\end{centering}

\caption{\label{fig:exec} Snapshot from 2-D PIC simulation of laser-driven plasma accelerator showing the wakefield from a laser propagating from left to right in a plasma channel: (top) transverse focusing (red) and defocusing (blue) fields; (bottom) accelerating (blue) and decelerating (red) fields. With appropriate phasing of the injection of the electron beam (green), approximately half of a period is available for simultaneous guiding and acceleration, immediately followed by simultaneous guiding and deceleration of the electron beam.}

\end{figure}
 
 Fig. \ref{fig:exec} shows the transverse and longitudinal electric fields in the wake produced by a laser in the quasi-linear regime ($a_0=1$). The alternating focusing-defocusing and accelerating-decelerating periods are shifted by $\sim \pi/2$, and approximately half a period is available for focusing.  This focusing phase is then split between accelerating or decelerating regions for the electron beam. The group velocity of the laser in the plasma is typically smaller than that of the electron beam, such that slippage occurs. Hence, an electron beam injected appropriately will be focused and accelerated to very high energy in a very short distance, then will slip ahead in the wake and reach the phase where it is efficiently decelerated while still being focused. This phasing effect has been studied in depth to enable extraction of the electron beam from the LPA at peak energy~\cite{Esarey2009, Leemans2006, Geddes2004, Mangles2004, Faure2004}.  By continuing the plasma beyond the length at which the beam achieves peak energy, acceleration and deceleration are accomplished in the same compact (cm-scale) structure.   For moderate energy spreads, photon production can then be conducted in the same plasma, at the phase interval between accelerating and decelerating regions where the longitudinal field is near zero. 
 
As an example, we consider the acceleration of an electron beam up to the energy required for production of 6 MeV photons and its subsequent deceleration. The desired energy of the beam after acceleration was 0.5 GeV and relative energy spread at or below 2$\%$. The parameters of the simulation to achieve this were determined from previous simulations using  scaling laws that have been demonstrated over a very wide range of energies~\cite{Vay2011}.  These scaling laws allow predictive design of LPA stages over a wide range of energies, and show that parameters such as energy spread remain constant. 
A laser of wavelength $\lambda=0.8\mu m$, profile $a(r,z)=a_0\exp{\left[-r^2/w^2\right]}\sin{\left[\pi z/L\right]}$ with a waist $w\sim25\mu$m, length $L\sim 28\mu$m and amplitude $a_0=1$, was injected in a plasma column of density $n_0=1.3\times10^{18}$ cm$^{-3}$ on axis with a parabolic transverse profile which provides laser guiding. 
An electron beam was injected in the wake with charge $Q=-10 $pC, energy $E\sim27$ MeV and relative energy spread $\Delta E/E\sim 0.1$, a Gaussian profile with r.m.s. width $\sigma_x=\sigma_y\sim0.56\mu$m and length $\sigma_z\sim0.5\mu$m and a normalized r.m.s. emittance $\epsilon_x=\epsilon_y\sim33\mu$m$\cdot$mrad. These are consistent with LPA injector parameters measured experimentally and reported elsewhere \cite{Gonsalves2011}. The beam was injected into the second plasma oscillation at the phase for acceleration and guiding.  This phase corresponded to $D\sim50\mu$m behind the peak of the laser pulse, or $D\sim1.7\lambda_p$.

\begin{figure}
\begin{centering}
\includegraphics[scale=0.3]{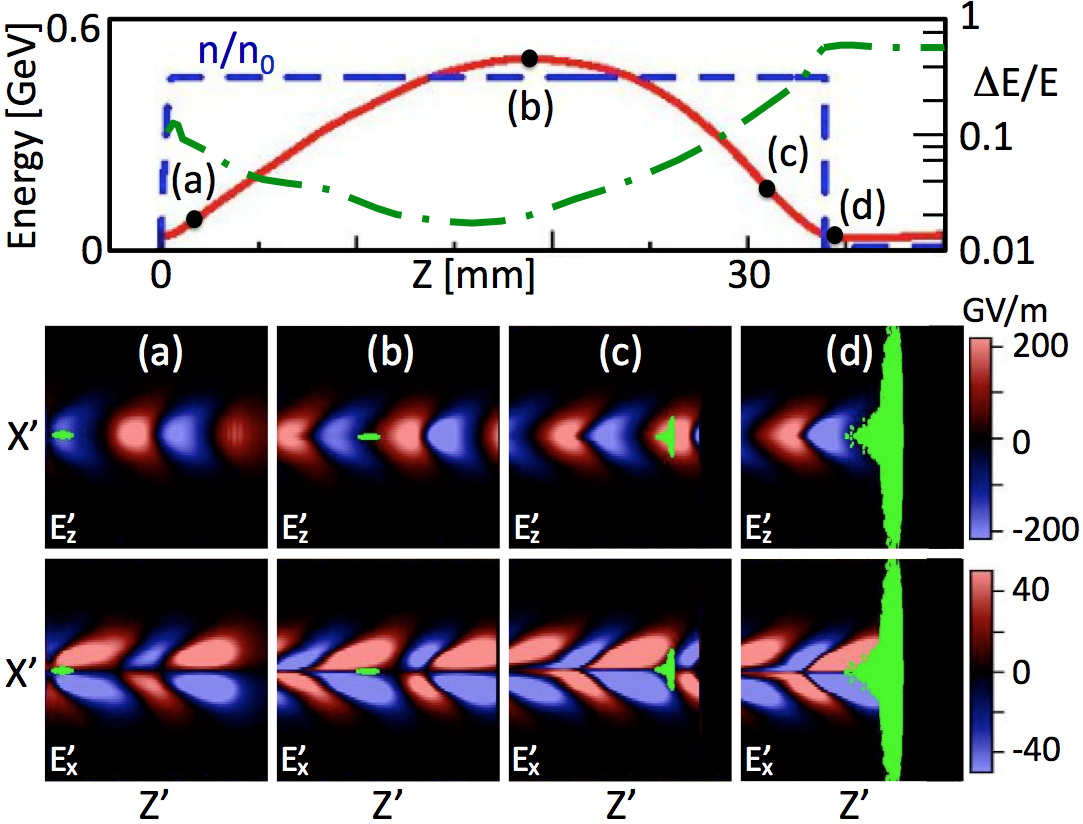}
\par\end{centering}

\caption{\label{fig:accel_decel_snapshots} (top) plasma profile on axis (dash blue), electron beam average energy (solid red) and energy spread histories (dot-dash green) in the laboratory frame; (bottom) snapshots of the transverse and longitudinal electric fields and electron beams (green) taken at propagation distances (a), (b) (c) and (d) indicated on top plot, in the Lorentz boosted simulation frame.}

\end{figure}

Fig. \ref{fig:accel_decel_snapshots} shows the plasma profile on axis, electron beam average energy and energy spread histories, as well as snapshots of the transverse and longitudinal electric fields and electron beams.
The snapshots are taken from a simulation using a Lorentz boosted frame~\cite{Vay2007} (boost at $\gamma_{boost}\sim 16.6$) and $Z'$, $E'_x$ and $E'_z$ are respectively the longitudinal coordinate, transverse electric field and longitudinal electric field in the simulation frame.
The electron beam is accelerated to $0.5$ GeV in $2$ cm, then decelerated to its injected energy in a slightly shorter distance of $~1.5$ cm, the asymmetry of the acceleration and deceleration distances being attributed to steepening of the wake structure from laser depletion \cite{Shadwick2009}. 
The relative energy spread falls from $10\%$ at injection to slightly below $2\%$ at peak energy (as prescribed for Thomson scattering), then rises steadily to nearly $100\%$ at the plasma exit at $z=34$mm. The beam energy is within 2 $\%$ of the peak value (0.5GeV) over the central 2 mm of propagation.  The length of this region is similar to the scattering pulse length required, as discussed further in Sec.~\ref{yield_section}.
Using the electron beam parameters from the Warp simulation, a simulation was performed with VDSR code~\cite{Chen2013} to compute the spectrum of photons from Thomson scattering giving $\approx13\%$ FWHM photon energy spread at 6 MeV, assuming a counter-propagating laser of wavelength $\lambda=0.8$~$\mu$m, amplitude $a_0=0.05$, length $27\lambda$ and focus width of $20\lambda$ consistent with eq.~(\ref{requirements_bandwidth_eq}). It has been previously demonstrated that the LPA energy and beam performance scale predictably with plasma density~\cite{Vay2011}. This allows other photon energies (e.g. 1 to 15 MeV) to be achieved with similar beam parameters to the present example.
Snapshots of the fields and electron beam at various times show that the beam is well focused when it enters the decelerating region of the wake but reaches the defocusing region toward the end of its deceleration, causing the beam to spread significantly transversely. The transverse spread induces non-uniform deceleration that results in higher energy spread and limits deceleration efficiency. Through parametric exploration that was enabled by the fast turnaround of Lorentz boosted frame simulations, the plasma length was adjusted to $\sim34$mm, which maximized averaged deceleration and minimized energy spread for the chosen laser, plasma and electron beam injection phase parameters.

\begin{figure}
\begin{centering}
\includegraphics[scale=0.4]{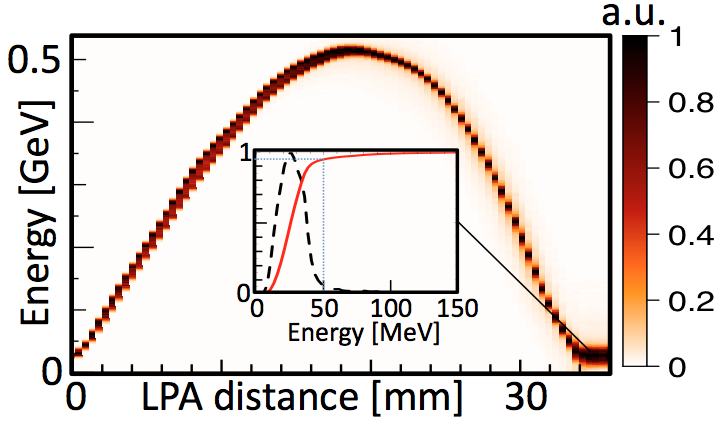}
\par\end{centering}

\caption{\label{fig:accel_decel}  Energy profile of the electron beam versus propagated distance in the plasma channel (in the laboratory frame); (insert) beam energy distribution (black dash) and cumulative energy distribution (solid red at exit).}

\end{figure}

Fig. \ref{fig:accel_decel} shows the energy profile of the electron beam versus propagated distance, as well as the beam energy distribution at exit, demonstrating efficient deceleration with $95\%$ of the electron beam energy below $50$ MeV.  This constitutes deceleration to less than 10$\%$ of the maximum energy, demonstrating the potential for efficient acceleration and deceleration over very short distances with LPAs. With regards to deceleration, further improvements are possible through the use of e.g., longitudinal plasma shaping and the addition of a passive deceleration plasma region.

\section{Estimation of the total yield}\label{yield_section}
	Total photon yield of the source can be found using the cross-section formalism. In this section we derive the total yield expressions for the case of interaction in vacuum, plasma waveguide to avoid laser pulse diffraction and plasma channel to avoid the diffraction of both the laser and electron beams. In the case when the recoil effect on the electron can be neglected one can use the Thomson cross-section given by $\sigma_T=\frac{8\pi}{3}r_e^2$, where $r_e=e^2/mc^2$ is classical electron radius. Total number of scattered photons is then given by \cite{LandauVol2, Villa1996}
\begin{equation}
N_\gamma=\sigma_T\int\limits_{-\infty}^{+\infty} v_{rel} n_e(t,\mathbf{r}) n_p(t,\mathbf{r})d^3\mathbf{r}dt\mathrm{,}
\label{total_yield_begin}
\end{equation}
where $n_e$ and $n_p$ are time-dependent densities of electrons and laser photons respectively and $v_{rel}$ is relative velocity of electrons and laser photons which can be approximated by $v_{rel}\approx 2c$.  In the case of round Gaussian bunches (see Fig.~\ref{schematics_figure}), electron and laser photon densities are given by
\begin{eqnarray}
n_{e}=\frac{N_{e}}{(2\pi)^{3/2}\sigma_{\perp,e}^2(z)\sigma_{l,e}}e^{\left(-\frac{\mathbf{r}_\perp^2}{2\sigma_{\perp,e}^2(z)}-\frac{(z-ct)^2}{2\sigma_{l,e}^2}\right)}\label{electron_dens_eq}\\
n_{p}=\frac{N_{p}}{(2\pi)^{3/2}\sigma_{\perp,p}^2(z)\sigma_{l,p}}e^{\left(-\frac{(\mathbf{r}_\perp-\Delta\mathbf{R})^2}{2\sigma_{\perp,p}^2(z)}-\frac{(z+ct-\Delta \zeta)^2}{2\sigma_{l,p}^2}\right)}\label{photon_dens_eq}\mathrm{,}
\end{eqnarray}
with subscripts $e$ and $p$ denoting electrons and laser photons respectively, $N$ denoting the total number of particles of a certain kind, $\sigma_\perp$ and $\sigma_l$ the transverse and longitudinal sizes respectively, $\Delta \zeta$ to take into account the relative delay between the pulses, and $\Delta\mathbf{R}$ for the transverse displacement of the bunches to take into account transverse jitter. Transverse sizes are in turn given by
\begin{eqnarray}
\sigma_{\perp,e}^2=\sigma_{e,0}^2\left(1+\frac{z^2}{\beta^{\star 2}_e} \right)\label{sigma_perp_e_eq}\\
\sigma_{\perp,p}^2=\sigma_{p,0}^2\left(1+\frac{\left(z-\Delta Z\right)^2}{\beta^{\star 2}_p} \right)\label{sigma_perp_p_eq}\mathrm{,}
\end{eqnarray}
where $\sigma_0$ denotes the spotsize of the bunch at focal position, $\beta^\star$ is the beta-function of the beam and $\Delta Z$ is introduced to take into account different longitudinal positions of the beams. The beta function is given by $\beta^{\star}=\sigma_{0}^2/\varepsilon_t$, where $\varepsilon_t$ is the transverse geometrical emittance of the beam. For the laser photon beam, $\varepsilon_{t,p}=\lambdabar_L/2=\lambda_L/4\pi$, so that the beta function of the photon beam equals its Rayleigh length. Substituting eqns.~(\ref{electron_dens_eq}) and (\ref{photon_dens_eq}) into eq.~(\ref{total_yield_begin}) one obtains after integrating over transverse coordinates and time
\begin{equation}
N_\gamma=\frac{2\sigma_T N_e N_p}{\left(2\pi \right)^{3/2}\sigma_l}\int\limits_{-\infty}^{\infty}\frac{e^{-\left(\frac{\Delta R^2}{\sigma_{\perp,e}^2+\sigma_{\perp,p}^2}+\frac{2\left(z-\Delta\zeta \right)^2}{\sigma_l^2}\right)}}{\sigma_{\perp,e}^2+\sigma_{\perp,p}^2}dz\mathrm{,}
\label{yield_only_z_remains_eq}
\end{equation}
 where 
 \begin{equation}
 \sigma_l=\sqrt{\sigma_{l,e}^2+\sigma_{l,p}^2}\mathrm{,}\label{sigma_l_eq}
 \end{equation}
 and $\sigma_{\perp,e}^2$ and $\sigma_{\perp,p}^2$ are given by eqns.~(\ref{sigma_perp_e_eq}) and (\ref{sigma_perp_p_eq}), respectively. Typical LPA electron beams are rather short (1-10 fs) compared to the scattering laser beams (several picoseconds), $\sigma_l$ can be replaced by the longitudinal size of the laser beam only. In general form, this integral needs to be taken numerically. However, several important analytical solutions exist and are presented below.
 
In the following we use the notation: quantities with subscript $\sigma_{FWHM}$ denote the FWHM values and quantities with subscript $\sigma_{\mu\mathrm{m}}$ are given in micrometers. For example, $\sigma_{e,0}$ is the root mean square transverse size of the electron beam in focus given in centimeters, $\sigma_{e,0,FWHM}$ is FWHM transverse size of the electron beam in focus in centimeters and $\sigma_{e,0,FWHM,\mu \mathrm{m}}$ is the FWHM transverse size of the electron beam given in micrometers. If not stated otherwise, the laser energy $E_L$ is given in Joules.
 
\subsection{Interaction in vacuum}\label{yield_vacuum_section}
Conventionally TS sources are operated in vacuum, where both laser and electron beams are diverging. In the case when electron and laser beams are interacting in vacuum and there are no relative displacements ($\Delta \zeta=0$, $\Delta\mathbf{R}=0$, $\Delta Z=0$), the integral in eq.~(\ref{yield_only_z_remains_eq}) can be analytically evaluated taking into account different beam sizes ($\sigma_{e,0}$ and $\sigma_{p,0}$) and beta functions ($\beta^\star_{e}$ and $\beta^\star_p$). The result of integration reads
\begin{equation}
N_\gamma=\frac{\sigma_T N_e N_p F(x)}{\sqrt{2\pi}\sigma_l\sqrt{\sigma_{e,0}^2+\sigma_{p,0}^2}}\frac{1}{\sqrt{\frac{\sigma_{e,0}^2}{\beta^{\star 2}_e}+\frac{\sigma_{p,0}^2}{\beta^{\star 2}_p}}}\mathrm{,}
\label{yield_mismatched_eq}
\end{equation}
with $x$ given by
\begin{equation}
x=\frac{\sqrt{2}}{\sigma_l}\sqrt{\frac{\sigma_{e,0}^2+\sigma_{p,0}^2}{\frac{\sigma_{e,0}^2}{\beta^{\star 2}_e}+\frac{\sigma_{p,0}^2}{\beta^{\star 2}_p}}}\mathrm{,}\label{x_equation}
\end{equation}
and $F(x)$ given by
\begin{equation}
\mathrm{F}(x)=e^{x^2}\left[1-\mathrm{erf}\left(x \right)\right]\mathrm{.}
\label{F_with_error_function_eq}
\end{equation}

The photon yield increases as a given laser beam is focused tightly because the laser amplitude rises. However, as derived in Sec.~\ref{spectral_shape_intensity_section} such intensity increase also causes broadening of the spectrum. Hence, the source bandwidth sets the upper limit on laser amplitude, and best yield is obtained running at this condition.
Fixing the laser pulse energy $E_L$ and its amplitude $a_{0,max}$ according to the source requirement condition of eq.~(\ref{requirements_bandwidth_eq}), one can find the connection between longitudinal and transverse sizes
\begin{equation}
\sigma_l=\frac{cE_L}{\left(2\pi \right)^{3/2}I_{max}\sigma_0^2}\mathrm{,}
\label{sigma_l_eq}
\end{equation}
with 
\begin{equation}
I_{max}=\frac{a_{0,max}^2\cdot 1.37\cdot10^{18}[\mathrm{W/cm}^2]}{\lambda_{L,\mu m}^2}\mathrm{,}\label{i_max_eq}
\end{equation}
where $\lambda_{L,\mu m}$ is the laser pulse wavelength measured in microns. Combining eqns.~(\ref{sigma_l_eq}) and (\ref{i_max_eq}) together with eqns.~(\ref{yield_mismatched_eq}) and (\ref{x_equation}) and using convenient units one can obtain the following formulas
\begin{equation}
\frac{N_\gamma}{N_e}=1.2\cdot\frac{\sigma_{p,0,\mu m}^2 \cdot a_{0,max}^2\cdot F(x)}{\lambda_{L,\mu m}^2}\cdot\frac{1}{\sqrt{1+\frac{\sigma_{e,0}^2}{\sigma_{p,0}^2}}}\cdot\frac{1}{\sqrt{1+\frac{\beta^{\star 2}_p\sigma_{e,0}^2}{\beta^{\star 2}_e\sigma_{p,0}^2}}}\mathrm{,}\label{n_gamma_full_eq}
\end{equation}
and
\begin{equation}
x=1.28\cdot 10^{-2}\cdot \frac{\sigma_{p,0,\mu m}^4\cdot a_{0,max}^2}{E_L \lambda_{L,\mu m}^3}\cdot \sqrt{\frac{1+\frac{\sigma_{e,0}^2}{\sigma_{p,0}^2}}{1+\frac{\beta^{\star 2}_p\sigma_{e,0}^2}{\beta^{\star 2}_e\sigma_{p,0}^2}}}\mathrm{.}\label{x_full_eq}
\end{equation}
Using equations (\ref{n_gamma_full_eq}) and (\ref{x_full_eq}) one can easily find the total photon yield in the case when laser and electron beams have the same focus position and diverge in vacuum with given beta-functions $\beta^\star_e$ and $\beta^\star_p$.
\begin{figure}
   \centering
   \subfigure{\includegraphics[width=0.4\textwidth]{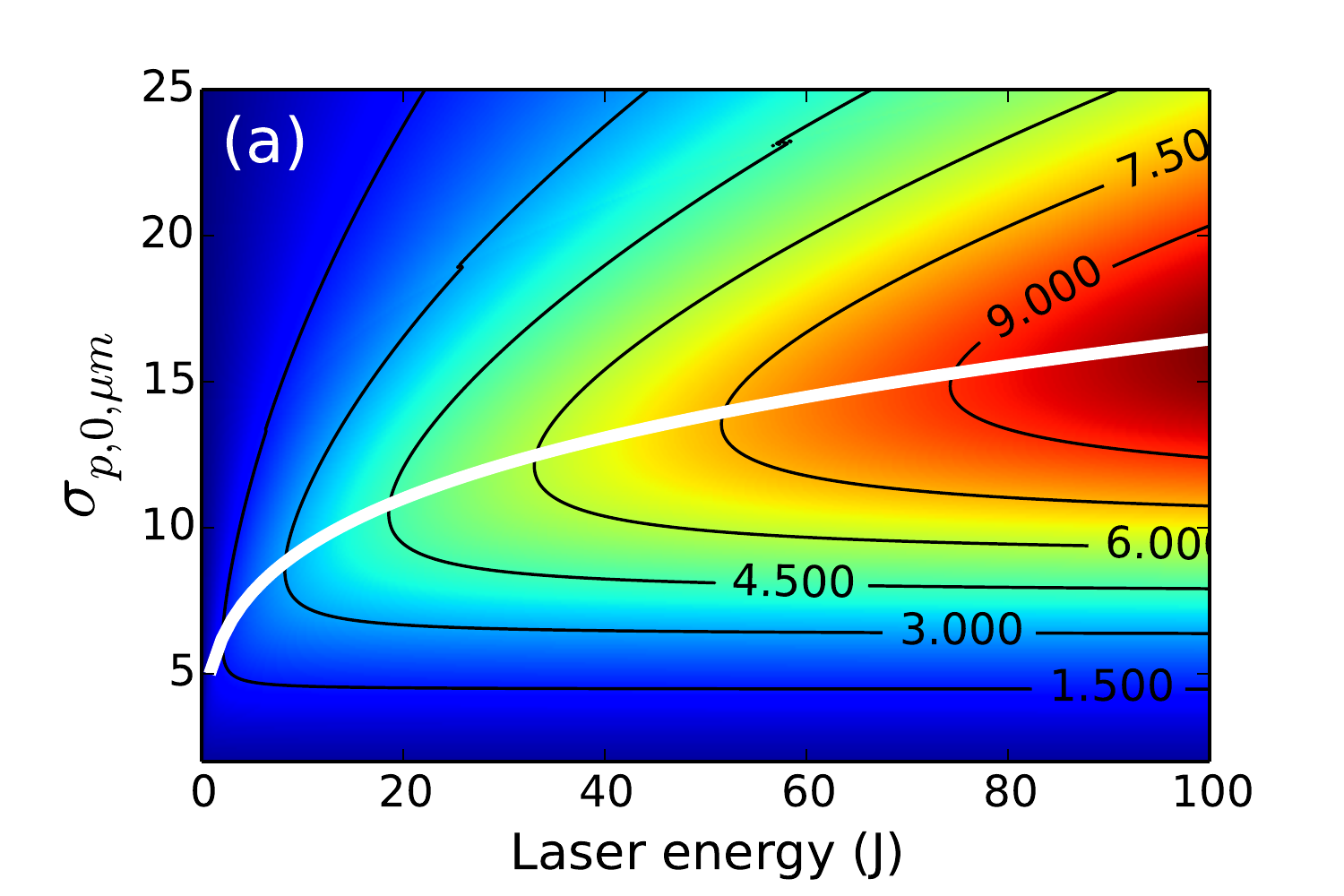}}
   \subfigure{\includegraphics[width=0.4\textwidth]{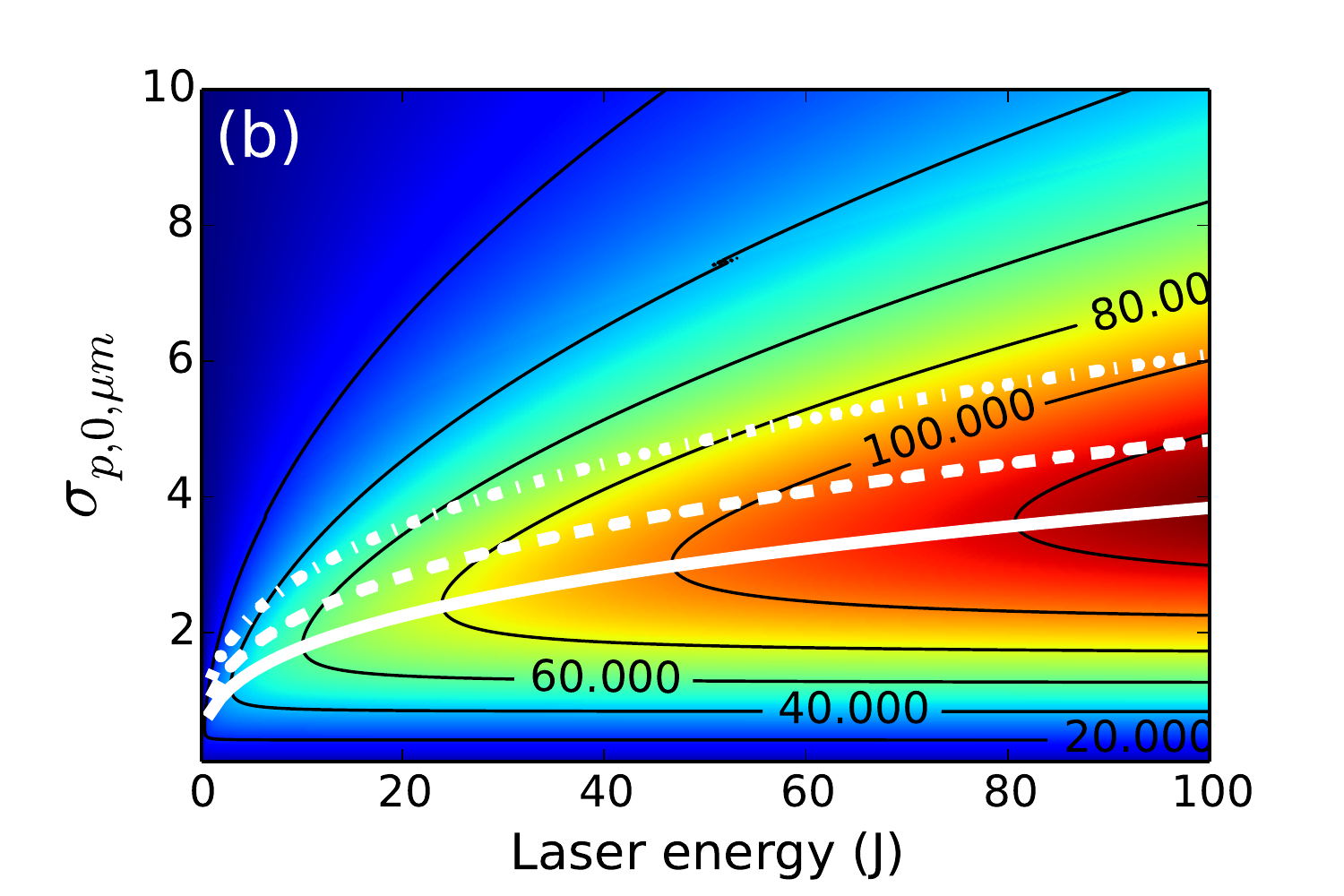}}
     \caption{(a): Total photon yield normalized to the number of electrons $N_\gamma/N_e$ (color-coded image) as a function of both
     the laser pulse energy in Joules (horizontal axis) and laser pulse spot size $\sigma_{p,0,\mu m}$ (left axis) for interaction in vacuum. Several isocontour lines are plotted in black color. Optimal
     transverse spot size $\sigma_{p, opt,\mu m}$ as a function of laser pulse energy and given by eq.~(\ref{sigma_opt_eq}) is shown with thick white line. (b): Same as on (a) , but for the case of interaction in a waveguide. The optimal transverse spot size is in this case given by eq.~(\ref{sigma_opt_waveguide_eq}) and is plotted with thick white line. Dashed and dot-dashed lines represent different fits to the optimum spot size and are explained in text.}\label{yield_intensity_included_figure}
\end{figure}

Typical emittance of the LPA electron beams is approximately 3 orders of magnitude smaller than the emittance of 1$\mu$m laser light.  Moreover, the transverse size of the LPA electron beams to be used for narrow-bandwidth TS sources is on the order of 1$\mu$m and thus much smaller than typical laser pulse spot size (10s of microns) needed for optimal photon yield, as discussed below. Considering this, one can neglect the terms containing $\sigma_e$ in eqns.~(\ref{n_gamma_full_eq}) and (\ref{x_full_eq}), yielding the simplified expressions
\begin{equation}
\frac{N_\gamma}{N_e}=1.2\cdot\frac{\sigma_{p,0,\mu m}^2 \cdot a_{0,max}^2\cdot F(x)}{\lambda_{L,\mu m}^2}\mathrm{,}\label{n_gamma_small_eq}
\end{equation}
and
\begin{equation}
x=1.28\cdot 10^{-2}\cdot \frac{\sigma_{p,0,\mu m}^4\cdot a_{0,max}^2}{E_L \lambda_{L,\mu m}^3}\mathrm{.}\label{x_small_eq}
\end{equation}

Yield calculations for the case of $a_{0,max}=0.2$ (so that broadening due to nonlinearity is on the order of 2 percent) are presented in Fig.~\ref{yield_intensity_included_figure}, where $N_\gamma/N_e$ is shown as a function of both the laser pulse energy and RMS laser pulse spot size $\sigma_{p,0,\mu m}$. For every laser pulse energy there exists an optimal laser pulse spot size and duration, which, according to calculations and geometrical considerations, can be found from the following condition
\begin{equation}
\sigma_l\approx 2\beta^{*}_p\mathrm{,}\label{sigma_l_rayleigh_eq}
\end{equation} 
or, in other words, the optimum laser pulse longitudinal size is approximately twice the Rayleigh range. The optimum laser pulse spot size as a function of energy then reads
\begin{equation}
\sigma_{p, opt,\mu m}\approx \frac{2.75\lambda_{L,\mu m}^{3/4}}{\sqrt{a_{0,max}}}E_L^{1/4}\mathrm{.}\label{sigma_opt_eq}
\end{equation}
This function is plotted on Fig.~\ref{yield_intensity_included_figure}~(a) with white curve. Using eqns.~(\ref{n_gamma_small_eq}) and (\ref{x_small_eq}) and taking into account the expression for the optimum laser pulse spot size, one can obtain the total photon yield per electron for the optimum laser pulse spot size and duration given by eq.~(\ref{sigma_l_rayleigh_eq})
\begin{equation}
\frac{N_\gamma}{N_e}\approx4.7 \cdot a_{0,max}\cdot \sqrt{\frac{E_L}{\lambda_{L,\mu m}}}\mathrm{.}\label{n_gamma_simple_eq}
\end{equation}
This formula agrees well with the results presented on Fig.~\ref{yield_intensity_included_figure}~(a).
Equation~(\ref{n_gamma_simple_eq}) indicates that using shorter wavelength laser sources might be beneficial provided one can control the laser pulse parameters (duration and spot size) to meet the optimum criteria. As an example, one can consider two cases: 1) one micron wavelength laser pulse and 2) frequency doubled micron wavelength laser pulse with second harmonic generation efficiency of 50 percent. In both cases the yield in the optimum cases will be the same. In reality, efficiency of second harmonic generation is higher than 50 percent, so that using the second harmonic of 1 or 0.8 micron lasers is beneficial. Moreover, for generation of the same photon energy, the required electron energy is $\sqrt{2}$ times lower for frequency doubled laser pulse, making accelerator systems more compact. 

Equation~(\ref{n_gamma_simple_eq}) shows that in the case of the interaction in free space the yield in the optimum case scales as square root of energy. This means that in order to increase the total yield, for example, 2 times, the laser pulse energy must be increased 4 times. This can lead to large laser systems. As is discussed in the next section, using waveguides to avoid laser pulse diffraction can be beneficial for reducing the laser energy requirements.

\subsection{Interaction in a plasma channel waveguide}\label{yield_waveguide_section}
Yield is limited for interaction of a diffracting laser beam with pencil-like ($\sigma_{e,\theta}\ll\sigma_{p,\theta}$) non-diffracting LPA electron beam (last section) due to the fact that photons are lost and do not participate in the interaction as the laser beam diffracts. The optimum laser pulse duration was found to be approximately $\sigma_l\approx 2 \beta^{\star}_p$. It is reasonable to assume that the total photon yield will be higher if one prevents the laser diffraction, for example using a waveguide, such as a plasma channel. In a plasma channel the radial density profile can be parabolic, which can exactly guide a Gaussian laser pulse in the low intensity limit provided the depth of the channel is equal to a critical value~\cite{Esarey2009}. Durfee III and Milchberg~\cite{III1993} demonstrated plasma-based guiding of a 25$\mu m$ FWHM (or $\sigma_{p,\mu m}\approx 10$) transversely wide Gaussian laser pulse for a distance over 24 times the Rayleigh range. Later, guiding of a 7$\mu$m FWHM ($\sigma_{p,\mu m} \approx 3$) laser pulses over a distance of 10 times the Rayleigh range was experimentally accomplished~\cite{Geddes2005}.
Alternatively, long ($\sim$10 cm) plasma channels have been generated using capillary discharges.

In the case of interaction inside a plasma channel waveguide the laser pulse Rayleigh length goes to infinity ($\beta^\star_p\rightarrow\infty$). Taking into account the bandwidth restriction on the laser pulse amplitude $a_{0,max}$, one obtains the following expression for the total photon yield

\begin{equation}
\frac{N_\gamma}{N_e}=0.096\cdot\frac{\sigma_{p,0,\mu m}}{\sigma_{e,\theta}}\frac{a_{0,max}^2\cdot F(x)}{\lambda_{L,\mu m}\cdot\sqrt{1+\frac{\sigma_{e,0}^2}{\sigma_{p,0}^2}}}\mathrm{,}\label{yield_waveguide_eq}
\end{equation}
where $\sigma_{e,\theta}$ is the angular divergence of the electron beam measured in radians. The term $\frac{\sigma_{p,0,\mu m}}{\sigma_{e,\theta}}$ gives the distance after which initially point-like electron beam will reach a transverse size equal to the laser beam transverse size $\sigma_{p,0,\mu m}$. This term, hence, is the characteristic interaction distance. In eq.~(\ref{yield_waveguide_eq}), the function $F(x)$ is given by eq.~(\ref{F_with_error_function_eq}) with $x$ given by the following expression
\begin{equation}
x=\frac{\sqrt{2}\cdot \sigma_{p,0}}{\sigma_{l}\cdot\sigma_{e,\theta}}\cdot \sqrt{1+\frac{\sigma_{e,0}^2}{\sigma_{p,0}^2}}=0.001\cdot \frac{a_{0,max}^2\cdot \sigma_{p,0,\mu m}^3}{\sigma_{e,\theta}\cdot E_L \cdot \lambda_{L,\mu m}^2}\cdot \sqrt{1+\frac{\sigma_{e,0}^2}{\sigma_{p,0}^2}}
\mathrm{.}\label{x_waveguide_eq}
\end{equation}
Again, for realistic cases $\sigma_{p,0}>\lambda_L$ and initial electron beam size $\sigma_{e,0}$ can be neglected with good accuracy, further simplifying the expressions.

An example of yield calculation using eqns.~(\ref{yield_waveguide_eq}) and (\ref{x_waveguide_eq}) is presented on Fig.~\ref{yield_intensity_included_figure}~(b). The electron beam size $\sigma_{e,0}$ was assumed to be zero (for the same reason as in the case of interaction in vacuum) and divergence was assumed to be $\sigma_{e,\theta}=0.1\cdot 10^{-3}$~rad. This divergence is approximately 10 times lower than the divergence of the LPA electron beam inside the LPA~\cite{Plateau2012}. The need for lower electron beam divergence and methods for divergence reduction were discussed in Sec.~\ref{using_LPA_section}. The optimum spot size can be roughly found from geometrical considerations. It is clear that the laser pulse duration has to be proportional to the interaction length $L_{int}=\sigma_{p,0}/\sigma_{e,\theta}$ over which the electron beam diverges such that its size equals then exceeds that of the (guided) scattering laser.
One can see that for every laser pulse energy there is an optimum laser pulse spot size. Numerical calculations show that $\sigma_l\cong4\cdot\frac{\sigma_{p,0}}{\sigma_{e,\theta}}$ in the optimum case, and the optimum laser pulse spot size is
\begin{equation}
\sigma_{p,opt,\mu m}\approx 7\cdot \left(\frac{\lambda_{L,\mu m}}{a_{0,max}} \right)^{\frac{2}{3}}\cdot\sigma_{e,\theta}^{\frac{1}{3}}E_L^{\frac{1}{3}}\mathrm{.}\label{sigma_opt_waveguide_eq}
\end{equation}
The plot of the optimum laser pulse spot size as a function of laser pulse energy for the case of interaction inside a waveguide is presented on Fig.~\ref{yield_intensity_included_figure}~(b) with white solid line. Dashed and dot-dashed lines on the same plot show the plots for the laser pulse spot size found from condition $\sigma_l=2\cdot\frac{\sigma_{p,0}}{\sigma_{e,\theta}}$ and $\sigma_l=\frac{\sigma_{p,0}}{\sigma_{e,\theta}}$ respectively. Even though the laser pulse duration was changed 2 and 4 times compared to the optimum case, one can calculate that the yield goes down only roughly 20 percent, making the choice of pulse duration in experiments flexible. The total yield in the optimum case can be found from eq.~(\ref{yield_waveguide_eq}) and reads
\begin{equation}
\frac{N_\gamma}{N_e}\approx 0.5 \cdot \lambda_{L,\mu m}^{-\frac{1}{3}}\cdot\sigma_{e,\theta}^{-\frac{2}{3}}\cdot E_L^{\frac{1}{3}}\cdot a_{0,max}^{\frac{4}{3}}\mathrm{.}
\end{equation}

It is important to note that the total photon yield in the case of the interaction in a plasma channel waveguide is approximately an order of magnitude larger for the same laser pulse  energy compared to the vacuum case, thus providing an optimization strategy. For narrow bandwidth, however, one should avoid multiple scattering beyond the limits discussed in Sec.~\ref{spectral_shape_intensity_section}. In general, using the plasma channel waveguide provides much higher yield than in the case of the interaction in vacuum, thus one can produce same amount of photons using less laser pulse energy even if the interaction in a waveguide is not set to the optimal parameters. In experiments, it may be difficult to reach the optimum spot size for certain laser pulse energies. For example, for an electron beam divergence of $\sigma_\theta=0.1$mrad and laser pulse energy of 0.1~J, the optimal laser spot size is $0.1\mu$m, i.e. less than a wavelength. This can not be achieved in experiments.
However, due to the high yield in the case of the interaction in a plasma channel waveguide, it is possible to choose a non-optimal set of parameters while still obtaining strong benefit versus vacuum operation, making experiments in this regime quite flexible. An example is provided in Sec.~\ref{numerical_sims_section}. 

\subsection{Interaction in near-hollow plasma channel}

It was proposed to guide electron beams using near-hollow plasma channels~\cite{Schroeder2013a}. In this case it is theoretically possible to create such a channel that will guide both the electron beam and laser pulse. It will still be necessary to decrease the electron beam divergence depending on the desired photon source bandwidth (i.e. approximately 10 times for 2 percent bandwidth photon sources leading to matched density of approximately $n_e\approx 5\cdot 10^{14}$cm$^{-3}$). 

In the case when neither the laser beam nor the electron beam evolve, the total yield is given by
\begin{equation}
N_\gamma=\frac{\sigma_T N_e N_p}{2\pi\left(\sigma_{e,0}^2+\sigma_{p,0}^2 \right)}\mathrm{,}
\label{yield_plasmachannel_eq}
\end{equation}
or, rewriting in convenient units,
\begin{equation}
\frac{N_\gamma}{N_e}\approx 53\cdot \frac{E_L\cdot \lambda_{L,\mu m}}{\sigma_{e,0,\mu m}^2+\sigma_{p,0,\mu m}^2}\mathrm{.}
\label{yield_plasmachannel_convenient_eq}
\end{equation}
In both waveguide and plasma channel cases, one is basically limited by the ability to guide laser pulses with as small transverse spot size as possible.  In general, the total photon yield in the case of a waveguide or a plasma channel is much higher than in the case of the interaction in vacuum providing flexibility in the experiment. According to our estimations, usage of plasma channel for both guiding the laser pulse and electron beam does not provide a considerable advantage over using just a waveguide. One can calculate that the yield in the case of a 1~J laser pulse with $\sigma_{p,\mu m}=5$ for electron beam with parameters same as on Fig.~\ref{yield_intensity_included_figure}~(a) is approximately $N_\gamma/N_e\sim 2$ in both the case of a waveguide and a plasma channel. However, use of a plasma channel might be beneficial in the case of electron beams with high divergence in less demanding applications. In such a case, the comparison of the yield can be done numerically or analytically using the formulas provided in this paper. The scattering laser energy sensitively depends on the guided spot size which can be achieved, and a guide similar to \cite{Geddes2005} at $\sigma_{p,\mu m} \approx 3$ would enable use of a 0.36 J laser for $N_\gamma/N_e\sim 2$.  Developing and implementing such guides in a Thomson scatter setup, compatible with the LPA, is important.

In principle, taking into account that for narrow bandwidth photon sources one has to use electron beams with low divergence, the most straightforward experimental setup consists of an LPA providing the electron beam, PMQ lenses for refocusing of the beam and a waveguide (either hollow fiber or plasma channel) for guiding of the backscattering laser pulse. Plasma optics provide a path to higher performance as well as compatibility with electron beam deceleration. For a photon source with 20$\%$ bandwidth, no additional refocusing of an electron beam is required.

\section{Design and simulation of TS photon sources for NRF and photofission studies}\label{numerical_sims_section}

Results presented in Sec.~\ref{spectral_shape_section} and Sec.~\ref{yield_section} are useful in designing the photon source. Provided that one have chosen the parameters of electron and laser beams such that the bandwidth requirement given by eq.~(\ref{requirements_bandwidth_eq}) are met, one can estimate the number of photons in given bandwidth $\kappa$. Total number of generated photons (of all energies) can be found using results of Sec.~\ref{yield_section} and depends on the geometry of the interaction. For an estimate of the number of photons in a given bandwidth $\kappa$, one can use eq.~(\ref{rough_estimate_photon_eq}) and simply multiply the total yield by the relative bandwidth $\kappa$. This estimate is within 20$\%$ for the cases discussed further in comparison with numerical simulations using particle tracking code VDSR~\cite{Chen2013}. For better accuracy, one can use eq.~(\ref{angle_energy_spec_circ_eq}) in the case when electron beam energy spread contribution is negligible compared to contribution due to beam divergence or use numerical integration as discussed in Sec.~(\ref{spectral_shape_energy_spread_section}) for the case when both electron beam energy spread and divergence are contributing to the spectrum bandwidth.

Numerical simulations using particle tracking in given electro-magnetic fields can be used to calculate the radiation directly for realistic source designs using eq.~(\ref{d2Idomegas_eq}). Numerical integration of eq.~(\ref{d2Idomegas_eq}) can be used to check and extend the  analytic expressions in the previous sections, which have neglected several effects such as, for example, electron beam evolution (phase space ellipse rotation), Gaussian distributions, and finite laser pulse bandwidth.  For obtaining the numerical results in this section we have used the code VDSR \cite{Chen2013}. The code VDSR has been thoroughly benchmarked against known theoretical results for synchrotron, betatron, undulator radiation as well as Thomson scattering. We have also used it for comparison of the results of Sec.~\ref{yield_section}, where the optimum laser pulse spot size was derived given the laser pulse energy and intensity as well as electron beam size. The agreement was within 5 percent and was limited due to the coarse numerical scanning of parameters. Further in this section, we provide an example design study of a LPA based gamma source relevant for NRF studies of $^{235}\mathrm{U}$ and photofission, and characterize the accuracy of the analytically derived formulae.

\subsection{Simulations of a photon source for NRF studies of $^{235}$U}

NRF studies demand narrow bandwidth at, for example, around 1.7 MeV for  $^{235}$U which is of primary interest for nuclear nonproliferation. Analytic calculations above indicated it is feasible to use TS from LPA electron beams to generate approximately $10^7$~photons/shot in  2 percent bandwidth. In this section we provide an example of a design study of a gamma-source which takes into account realistic beam parameters and evolution.  

  In order to generate approximately 2 percent bandwidth, we begin with the following values for electron energy spread and divergence according to eq.~(\ref{requirements_bandwidth_eq}) $\frac{\sigma_{\gamma_e,FWHM}}{\gamma_e}=0.9\%$ and $\gamma_e\sigma_{\theta,FWHM}=0.05$. We take a backscattering laser pulse with $a_0=0.15$ so that broadening of the spectrum due to the nonlinearity is approximately 1 percent. Electron beam central energy of 270~MeV yields generated photons  energy around 1.7~MeV for $\lambda_{L,\mu \mathrm{m}}=0.8$. Taking into account the estimations for the relative number of photons in the bandwidth $\kappa=0.02$ given by eq.~(\ref{rough_estimate_photon_eq}) one can calculate that the total number of photons should be on the order of $N_\gamma\approx 5\cdot 10^8$. For a typical number of electrons in the LPA electron bunch of $N_e\approx 10^8$, each electron on average must scatter 5 times, hence $\frac{N_\gamma}{N_e}\approx 5$. Depending on the geometry of the interaction one can now find the needed laser pulse energy to achieve the total yield of $\frac{N_\gamma}{N_e}\approx 5$ using the results of Sec.~\ref{yield_section}. To do this, we consider the case of  interaction in vacuum and inside a waveguide.

\begin{figure}
  \centering
    \subfigure{\includegraphics[width=0.45\textwidth]{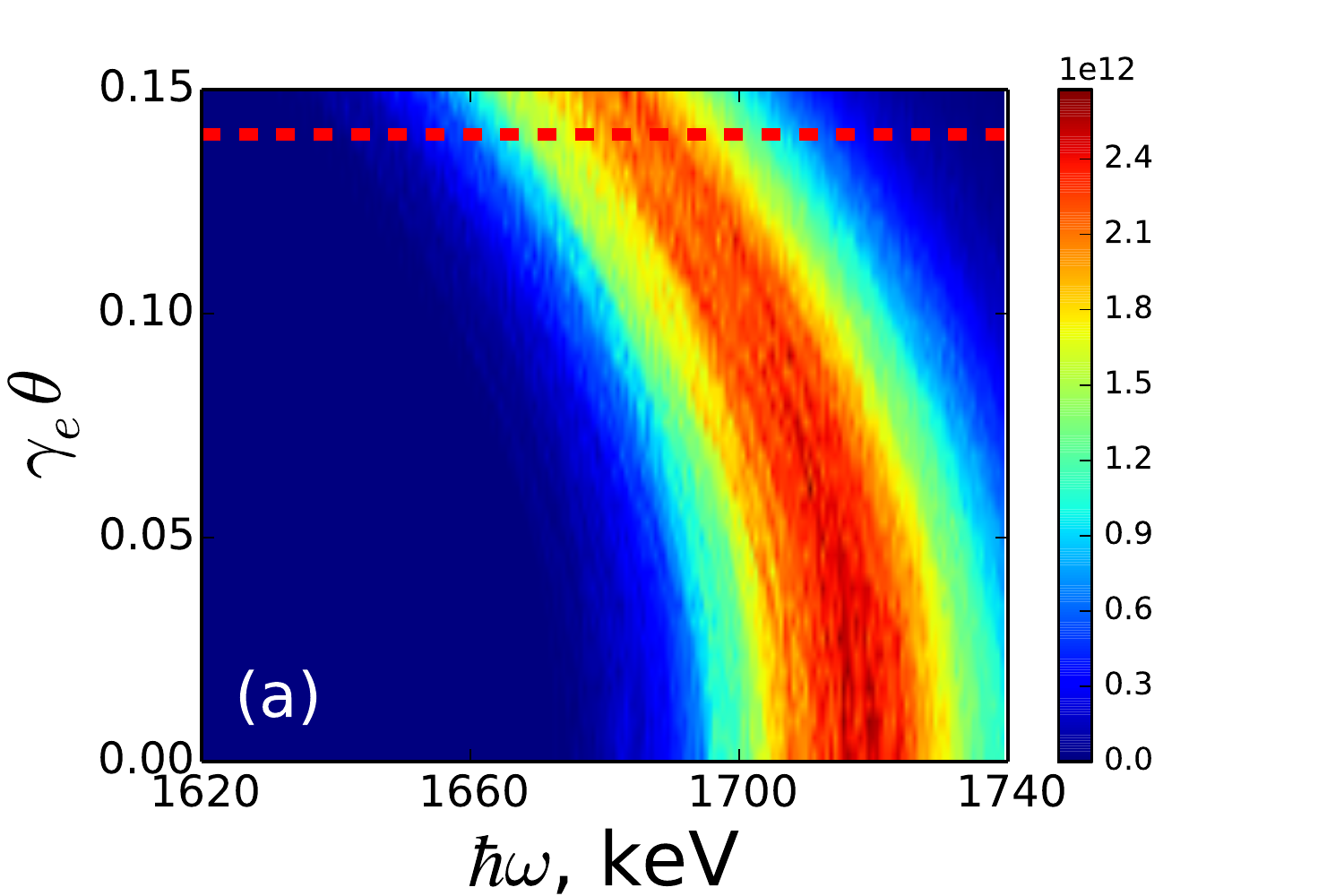}}
     \subfigure{\includegraphics[width=0.45\textwidth]{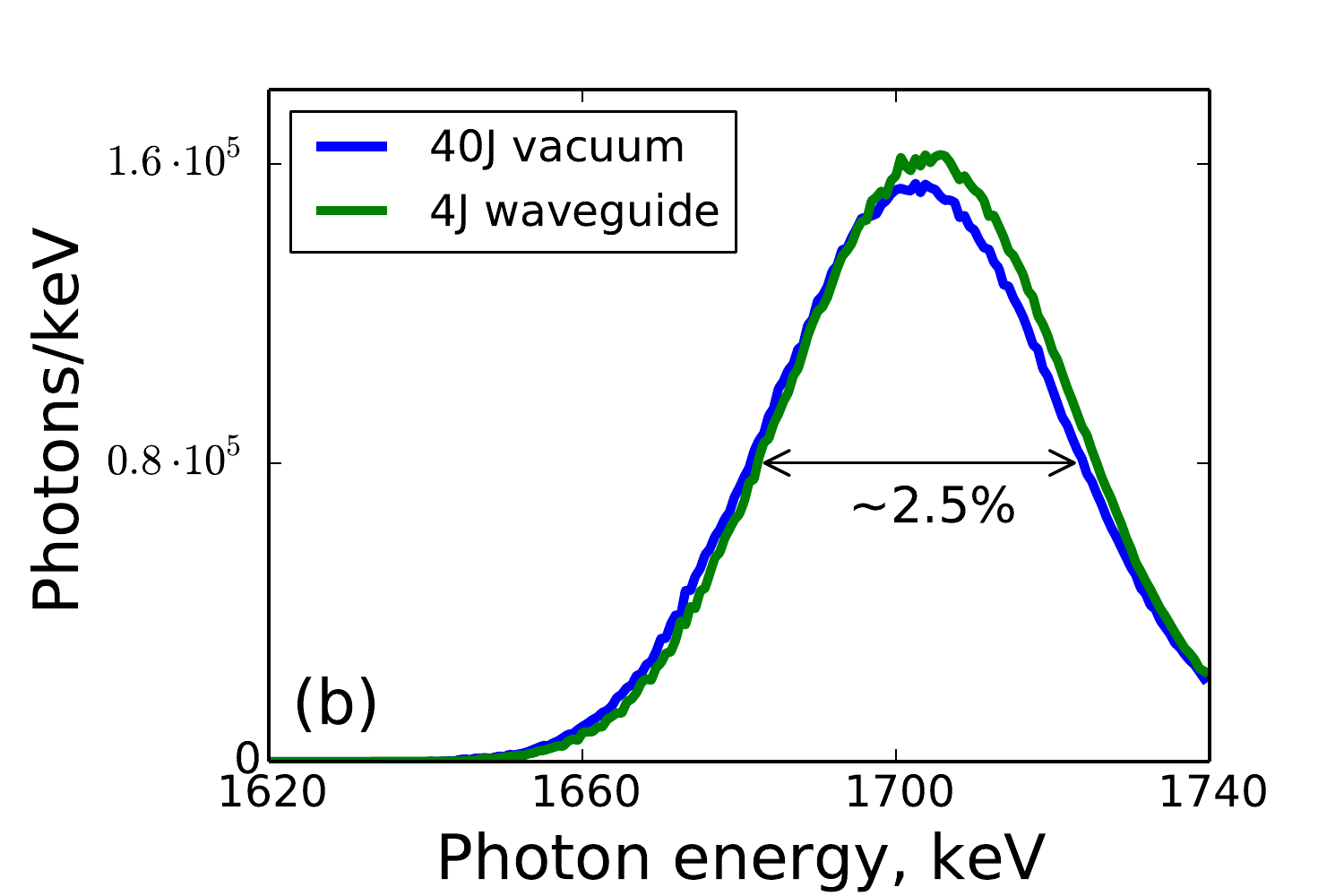}}
     \caption{(a): Photon energy and angular spectrum in photons/keV/sr as a function of photon energy in keV (longitudinal axis) and angle $\gamma_e\theta$ (vertical axis). Red dashed line represents the collimation angle calculated using eq.~(\ref{collimation_angle_eq}). (b): Plot of the number of photons per keV as a function of photon energy in keV obtained after the collimation in the case of the interaction of 40J pulse in vacuum (blue line) and 5J pulse in the waveguide (green line).}\label{photons_NRF_figure}
\end{figure}

The analytical calculations and estimations of Sec.~\ref{yield_vacuum_section} and Sec.~\ref{estimation_collimation_section} predict the  laser pulse energy eq.~(\ref{n_gamma_simple_eq}), optimum spot size eq.~(\ref{sigma_opt_eq}) and duration eq.~(\ref{sigma_l_eq}) for scattering in vacuum to produce  approximately $10^7$ photons in approximately 2$\%$ bandwidth. The corresponding numbers are $E_L=40$~J, $\sigma_{0,\mu m}=15$ and $\sigma_{l,\mu m}=8800$ respectively. 
The numerical results of VDSR calculations for this case are presented on Fig.~\ref{photons_NRF_figure}.  Figure~\ref{photons_NRF_figure}(a) shows the normalized photon spectrum in photons per keV per steradian as a function of photon energy in keV (longitudinal axis) and angle $\gamma_e\theta$ (vertical axis). 
To obtain the photon energy spectrum after collimation we have integrated the energy and angular spectrum in angle from $\gamma_e\theta=0$ to $\gamma_e\theta_c=0.14$ (outlined with the red dashed line on Fig.~\ref{photons_NRF_figure}) in accordance with eq.~(\ref{collimation_angle_eq}) and 
plotted the results on Fig.~\ref{photons_NRF_figure}(b) (blue color). Figure~\ref{photons_NRF_figure}(b) represents the number of photons per keV as a function of photon energy in keV. The FWHM bandwidth of the source is 
2.5\% and the total photon number obtained by integrating the function on Fig.~\ref{photons_NRF_figure}(b) in energy is $N_{\gamma ,0.02}= 1.1\cdot 10^7$. These are in good agreement with the analytical estimates. Varying the collimation angle one can obtain a photon source with slightly narrower bandwidth, but the photon yield then goes down. On the other hand, increasing the collimation angle can yield to more photons in broader bandwidth. The fine tuning of collimation angle is then dependent on the exact needs of the experiment.  In any case, the scattering laser energy is much higher than the Joule-class lasers needed to drive LPAs of this energy, motivating guiding.

Use of a waveguide to reduce scattering laser energy and bulk was next evaluated, starting from the formulae of Sec.~\ref{yield_waveguide_section}. For the parameters of the electron beam described above and in order to scatter on the average 5 times from each electron, the optimum laser pulse energy in the waveguide case is around 1J and the optimal spotsize is around 1$\mu$m.

 This is hard to achieve in experiments. We have hence chosen parameters that are not optimal, but still provide same amount of photons as in the case of interaction in the free space for much less laser pulse energy. Using results of Sec.~\ref{yield_waveguide_section} one can calculate that a laser pulse with energy 5J and spot size $\sigma_{p,0,\mu m}=6$ propagating in a waveguide will produce $N_\gamma=5N_e$ and thus one can expect photon source parameters to be same as in the case of the interaction in vacuum described in the previous section. This is indeed so, as one can see in Fig.~\ref{photons_NRF_figure} (b, green color). The photon energy spectrum looks very similar to the case of the interaction in vacuum and the total number of photons within 2.5$\%$ FWHM bandwidth is $N_{\gamma ,0.02}\approx 1.1\cdot 10^7$ in good agreement with analytical calculations. As mentioned above, the required laser pulse energy will go down depending on the ability to guide laser pulses with smaller spot sizes. For example, in the case when the guided spot size is  $\sigma_{p,0,\mu m}=5$ (compared to 6 used in the numerical simulations), the yield calculations show that the required laser pulse energy goes down to 3~J and for $\sigma_{p,0,\mu m}=3$, the required laser pulse energy is 1~J.


\subsection{Simulations of a photon source for photofission or radiography experiments}

For photofission and radiography applications, energy spreads at the 10-20$\%$ level are beneficial. The relaxed energy spread requirement allows TS sources to produce yields using simplified setups, and hence these are attractive first applications. 

The rough requirements for such a source are presented in the right column of Table~\ref{higs_table}. We have chosen electron beam energy to be 650~MeV, so that the generated photon energy is approximately 9~MeV for $\lambda_{L,\mu m}=0.8$. In order to generate $10^8$ photons in 10$\%$ FWHM bandwidth one needs the total number of photons to be approximately $N_\gamma=10^9$ according to the estimations provided in Sec.~\ref{estimation_collimation_section}. Taking into account that the typical number of electrons in LPA beam is approximately $10^8$, each electron on the average must scatter ten times, or, in other words, $\frac{N_\gamma}{N_e}\approx 10$. 

\begin{figure}
  \centering
    \subfigure{\includegraphics[width=0.45\textwidth]{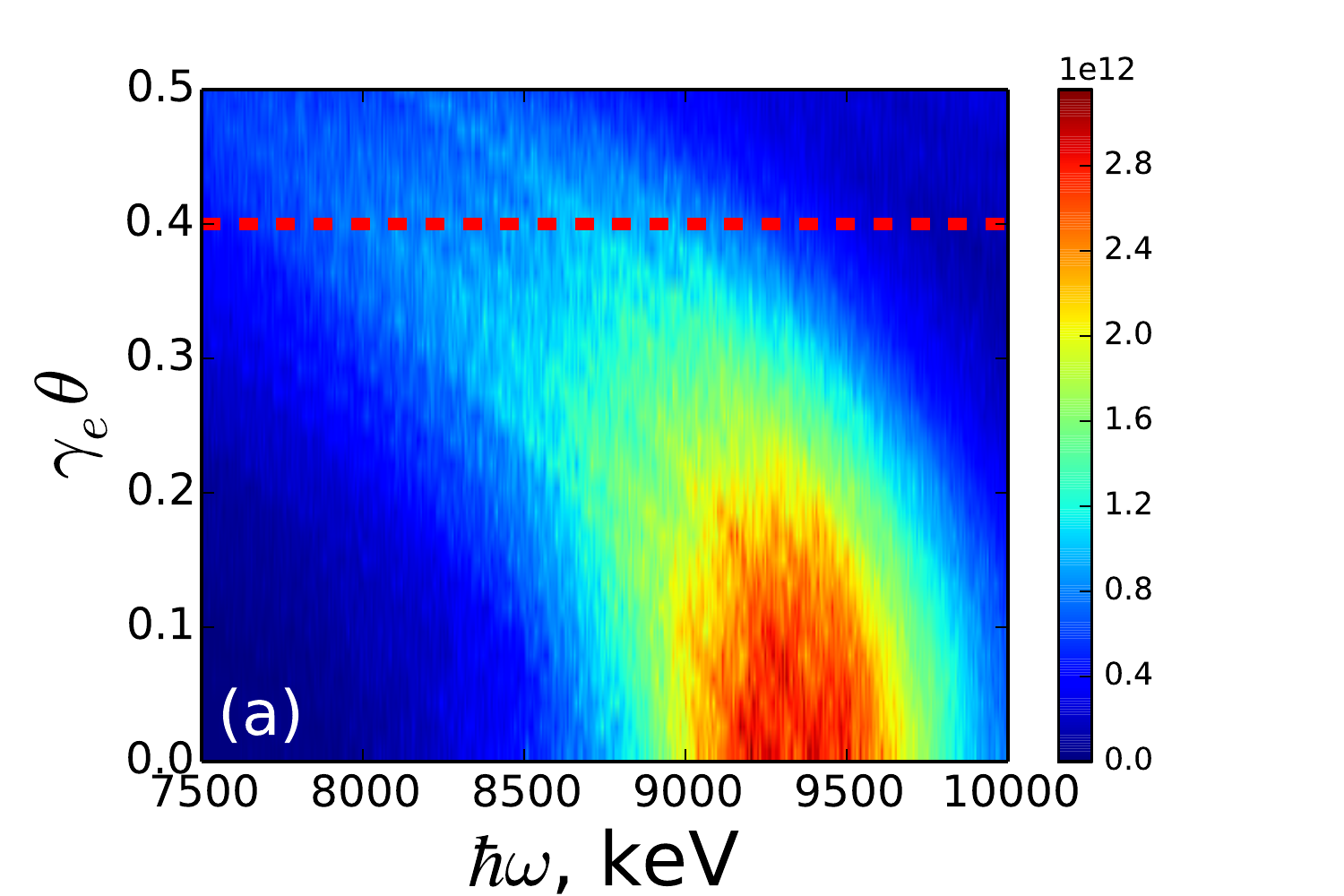}}
     \subfigure{\includegraphics[width=0.45\textwidth]{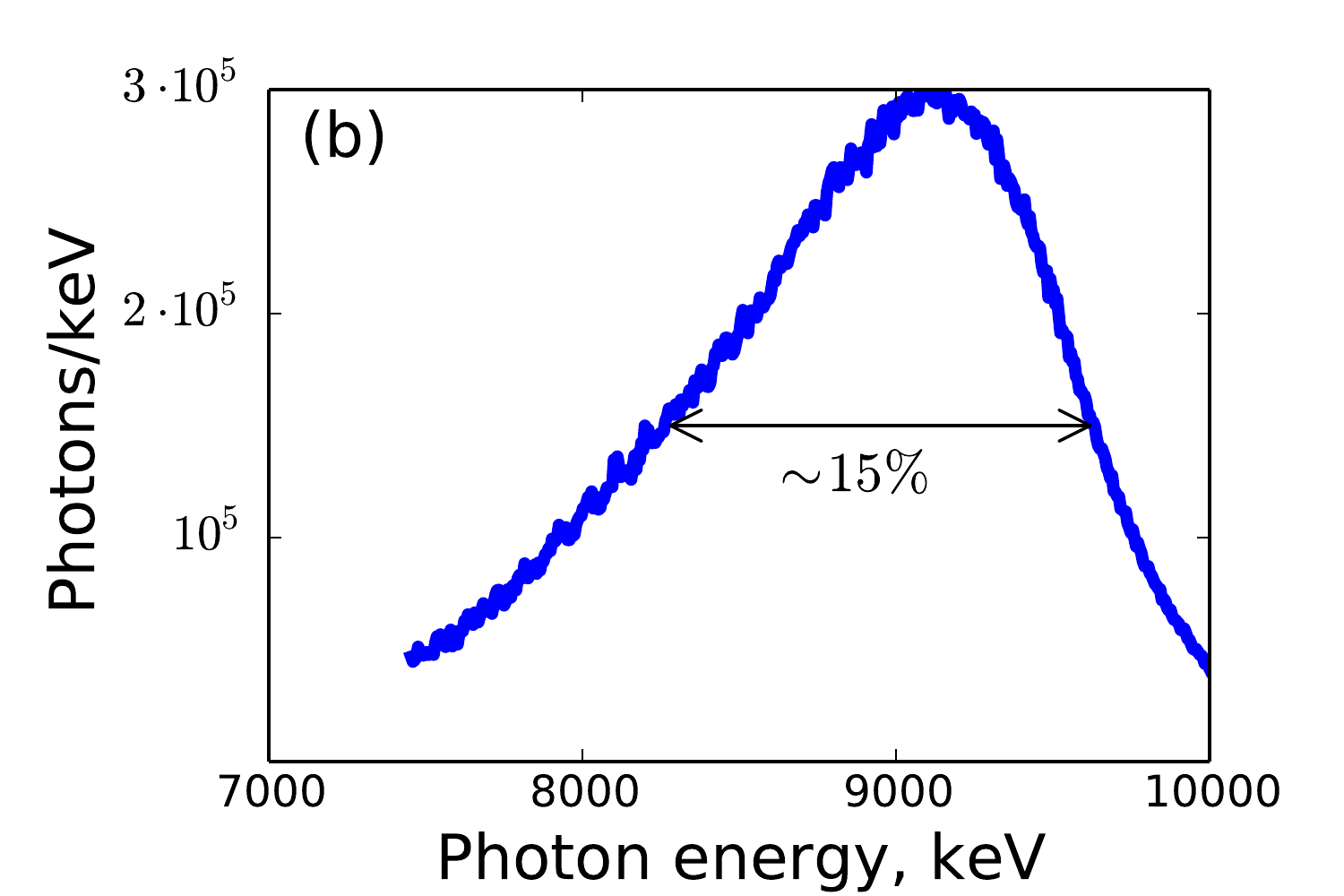}}
     \caption{(a): Photon energy and angular spectrum in arbitrary units as a function of photon energy in keV (longitudinal axis) and angle $\gamma_e\theta$ (vertical axis). (b): Number of photons per keV as a function of photon energy in keV obtained after the collimation with the maximum allowed angle $\gamma_e\theta=0.4$ (red dashed line on (a)) in the case of the interaction of 40J pulse in vacuum (blue line) and 4J pulse in the waveguide (green line).}\label{photons_PF_figure}
\end{figure}

The relaxed energy spread requirement allows efficient scattering without guiding of the scattering laser, because a higher laser amplitude can be used. We have chosen laser pulse amplitude to be $a_0=0.3$ so that the broadening due to nonlinear effects is approximately 4$\%$. Using eqns.~(\ref{n_gamma_simple_eq}),(\ref{sigma_opt_eq}) and (\ref{sigma_l_eq}) one can find that optimal unguided laser pulse energy is equal to $E_L=40$J, spot size is equal to $\sigma_{p,0,\mu m}=7.6$ and longitudinal size is equal to $\sigma_{l,\mu m}=2400$ respectively. The electron beam parameters were chosen to be $\frac{\sigma_{\gamma_e,FWHM}}{\gamma_e}=3\%$ and $\gamma_e\sigma_{\theta,FWHM}=0.4$ which is within reach for current LPA experiments. The total photon source FWHM bandwidth is predicted to be roughly 10$\%$ according to eq.~(\ref{requirements_bandwidth_eq}). Figure~\ref{photons_PF_figure} presents results of numerical simulations using the code VDSR using the parameters described above. Figure~\ref{photons_PF_figure}(a) shows the energy and angular spectrum given in photons per keV per  steradian. The collimation angle was chosen to be $\gamma_e\theta_c=0.4$ (red dashed line) and is very close to the angle calculated in accordance with eq.~(\ref{collimation_angle_eq}), which is equal to $0.3$. Photon energy spectrum (in same units as in Fig.~\ref{photons_NRF_figure}(b)) calculated by integrating the energy and angular spectrum in polar angle from 0 to $\theta_c$ is presented on Fig.~\ref{photons_PF_figure}(b). The photon source FWHM bandwidth is approximately 15 percent and the total number of photons in this bandwidth is approximately $0.8\cdot 10^8$ in fair agreement with the analytical predictions. It is worth noticing that in this case the required laser pulse energy (40~J) is rather high. Again, one can calculate using the results of this paper that in the case of guiding of a laser pulse with spot size of $\sigma_{p,\mu m}=5$ the energy of the laser pulse will go down to 5~J. Simulations show that estimations provided in this paper work better for small energy spread photon beams ($<10\%$).

\section{Discussion and conclusions}\label{conclusions_section}

In this paper we have
provided a detailed study of the photon sources
based on Thomson Scattering.  The main focus of the paper has been on
generation of the gamma-ray photons, but all results can be used also
in the case of the generation of X-rays.  Analytical calculations and
estimations presented in this paper can be used in designing photon
sources and optimizing experiments.  We have outlined the
contributions of electron beam energy spread, divergence, and laser
pulse intensity on the total source bandwidth.  In the case of the LPA
electron beams, for generation of several-percent-level gamma-sources,
it is necessary to reduce the electron beam divergence by
approximately an order of magnitude.  We have outlined possible
methods that include using permanent magnetic quadrupole lenses,
plasma lens, controlling the injection of electrons into the
accelerating structure, and a density downramp.  We have also
presented total yield calculations for different interaction
geometries and identified the optimum spot size for a given laser
energy and intensity.  The main limitation of the yield in the case of
interaction in vacuum is due to laser pulse diffraction and thus the
interaction distance is limited to approximately twice the Rayleigh
range.  Waveguides or plasma channels are beneficial as this
limitation is mitigated.  This is shown both analytically and
numerically.  Results of analytical calculations agree well with the
numerical simulations using the code VDSR. Examples of design studies
of LPA-based photon sources capable of performing the NRF studies of
$^{235}$U, as well as photofission studies were presented.  
In addition, the used of a plasma as a compact beam dump has been
studied. TS photon
sources from LPA electron beams are a promising path towards high
intensity femtosecond x- and gamma-ray sources and allow generation of
narrow-bandwidth photon spectra.  In theory, LPA-based TS sources can
compete with such large facilities as HIGS. Although the results were
focused on using the LPA electron beams, they can, without any
changes, be applied to conventional electron beams from linear
accelerators or storage rings.

\acknowledgements
This work was supported by the U.S. Dept. of Energy National Nuclear Security administration DNN/NA-22, and by the Office of Science Office of High Energy Physics,  under Contract No. DE-AC02-05CH11231.  We would like to acknowledge fruitful discussions with M.~Zolotorev, C.~Benedetti, M.~Chen, S.S.~Bulanov and F.~Rossi.

\appendix
\section{Pointing errors}\label{pointing_errors_section}
Electron and laser beams to be used in TS will have transverse size on the order of several microns or tens of microns. Transverse jitter and other errors are inevitably present in the laboratory. The question arises of whether such errors will be detrimental  for the photon source. To take into account the pointing errors, such as transverse and longitudinal jitter as well as electron and laser beam timings, in the general case, one has to numerically solve the integral in eq.~(\ref{yield_only_z_remains_eq}). As an example we consider the case of interaction in vacuum with matched beam sizes and beta functions. In such a case, the study of the influence of pointing errors is equivalent to studying the properties of the following integral
\begin{equation}
I(\Delta\tilde\zeta,\Delta  \tilde z,  \Delta \tilde R)=\int\limits_{-\infty}^{+\infty}\frac{e^{-2(z-\Delta\tilde\zeta)^2}}{f(z)}e^{-\frac{\Delta\tilde R^2}{f(z)}}dz\mathrm{,}
\label{integrand_error_eq}
\end{equation}
with normalized delay between pulses $\Delta\tilde\zeta=\Delta\zeta/\sigma_l$, normalized longitudinal $\Delta\tilde z=\Delta z/\sigma_l$ and transverse $\Delta \tilde R=\Delta R/\sigma_0$ pointing errors, and with $f(z)$ given by
\begin{equation}
f(z)=1+\frac{\left[z^2+(z-\Delta\tilde z)^2\right]}{x^2}\mathrm{.}
\end{equation}
For the optimal spot size and duration, $x=\sqrt{2}\beta^{\star}_p/\sigma_l$ can be calculated using eq.~(\ref{x_small_eq}) and the result yields $x_{opt}\approx 0.7$.
\begin{figure}
   \centering
    \includegraphics[width=0.5\textwidth]{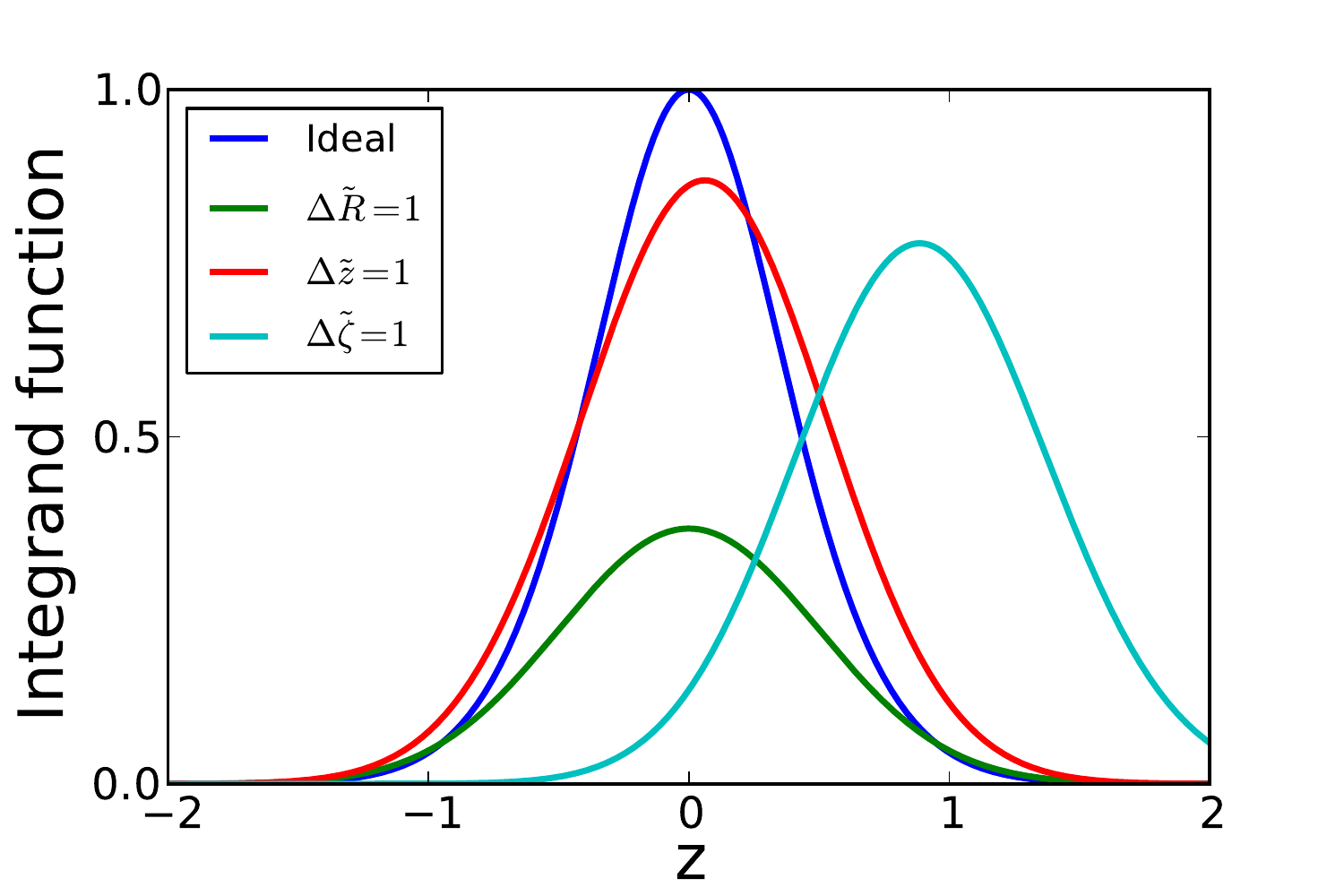}
     \caption{Integrand of the r.h.s. of eq.~(\ref{integrand_error_eq}) as a function of $z$ for 4 different cases: 1) ideal case (no errors, blue line); 2) case of non-zero transverse jitter ($\Delta \tilde R=1$, every other error is zero, green line); 3) case of non-zero longitudinal jitter ($\Delta \tilde z=1$, every other error is zero, red line); 4) case of the non-zero delay between pulses ($\Delta \tilde \zeta=1$, every other error is zero, cyan line). }\label{integrand_error_figure}
\end{figure}
\begin{figure}
   \centering
    \includegraphics[width=0.5\textwidth]{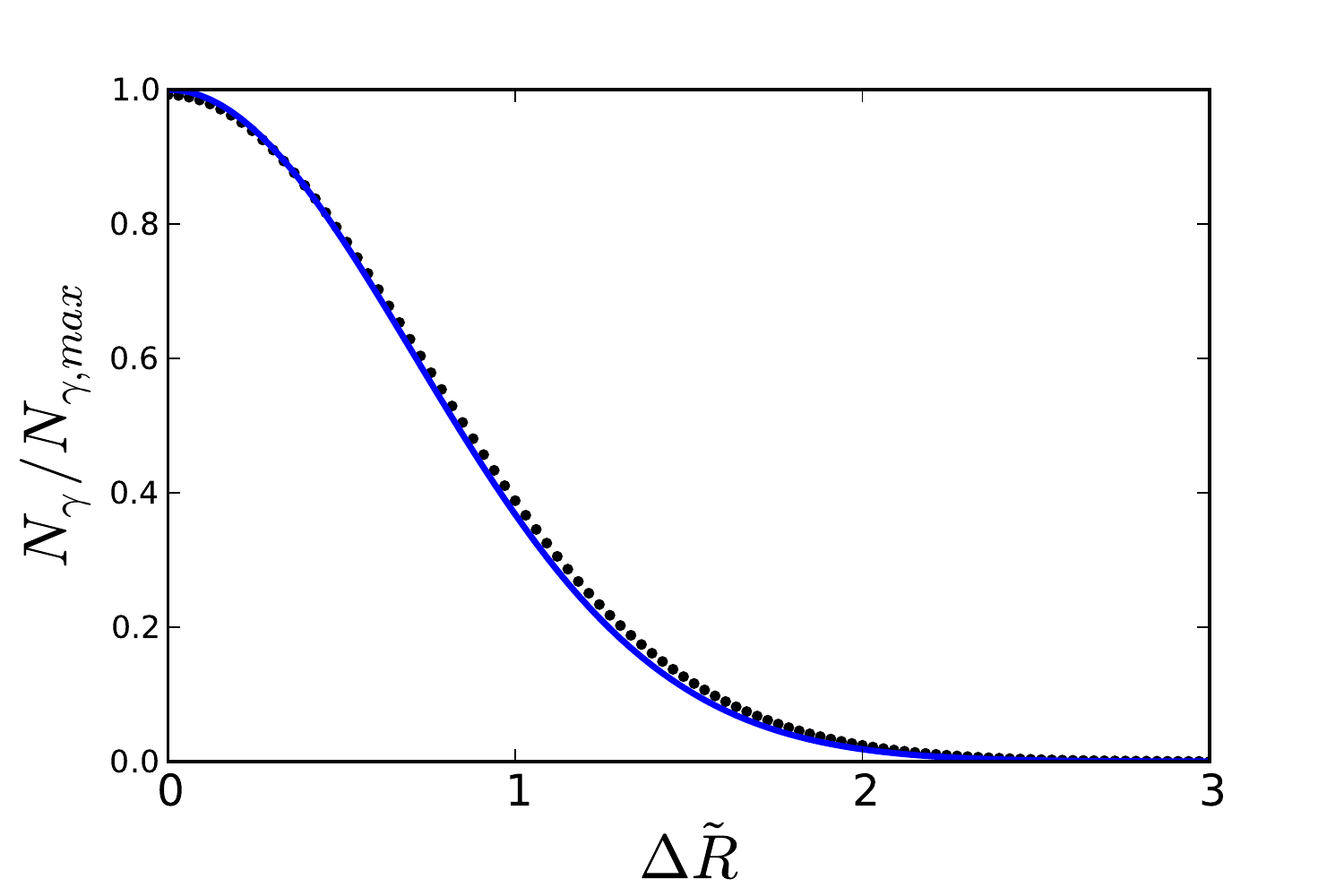}
     \caption{Total photon yield normalized to the ideal photon yield as a function of normalized transverse jitter $\Delta \tilde R$ calculated from eq.~(\ref{integrand_error_eq}) is shown with black circles. Blue solid line shows the fit $N=e^{-\Delta\tilde R^2}$.}\label{transverse_jitter_figure}
\end{figure}
Effects of the normalized delay and longitudinal and transverse pointing errors are summarized on Fig.~\ref{integrand_error_figure}, where the integrand of the right hand side of eq.~(\ref{integrand_error_eq}) as a function of $z$ is plotted for different cases: 1) ideal case (no errors, blue line); 2) case of non-zero transverse jitter ($\Delta \tilde R=1$, every other error is zero, green line); 3) case of non-zero longitudinal jitter ($\Delta \tilde z=1$, every other error is zero, red line); 4) case of the non-zero delay between pulses ( $\Delta \tilde \zeta=1$, every other error is zero, cyan line).  The main yield decrease in experiments will hence come from transverse jitter. The total photon yield normalized to the ideal photon yield is plotted on Fig.~\ref{transverse_jitter_figure} as a function of transverse jitter $\Delta \tilde R$. Transverse jitter has to be controlled on the order of  spot size. In the case when the transverse jitter is about half of the spot size ($\Delta \tilde R=0.5$), the yield goes down only by approximately 30 percent. Better control than half of the spot size has been demonstrated in experiments on colliding pulse laser injection~\cite{Geddes2011}.

Errors due to longitudinal focal jitter will be negligible as typical lasers have jitter of below 10$\%$ of Rayleigh length for which yield effects are at the 1 percent level. Similarly timing jitter effects  will be negligible because, typical pulse lengths of the scattering laser are 10~ps level while timing jitter control at the sub-ps level is routine, and 50~fs level has been demonstrated by splitting~\cite{Geddes2011} or path control~\cite{Wilcox2010}.

In the general case such non-ideal effects can be taken into account by numerical integration of eq.~(\ref{yield_only_z_remains_eq}) and will depend on exact experimental parameters. Current experimental capabilities allow generation of hard photons with total yield decrease of $<30$ percent compared to the optimum (no transverse jitter) case. Hence, using LPA electron beams for TS photon sources is quite reasonable.

\bibliography{library}

\end{document}